\documentclass[aps,prb,twocolumn,superscriptaddress,showpacs]{revtex4}
\usepackage{graphicx}
\usepackage{dcolumn}
\usepackage{epsfig}

\begin{document}

\newcommand{\etal}[0]{\textit{et al.}}
\newcommand{\alumina}[0]{Al$_2$O$_3$}
\newcommand{\Kalumina}[0]{$\kappa$-Al$_2$O$_3$}
\newcommand{\Aalumina}[0]{$\alpha$-Al$_2$O$_3$}
\newcommand{\Galumina}[0]{$\gamma$-Al$_2$O$_3$}
\newcommand{\bea}[0]{\begin{eqnarray}}
\newcommand{\eea}[0]{\end{eqnarray}}
\newcommand{\nn}[0]{\nonumber}
\newcommand{\mtext}[1]{\mbox{\tiny{#1}}}
\renewcommand{\vec}[1]{{\bf #1}}
\newcommand{\op}[1]{{\bf\hat{#1}}}
\newcommand{\galpha}[0]{$\alpha$}
\newcommand{\gbeta}[0]{$\beta$}
\newcommand{\ggamma}[0]{$\gamma$}
\newcommand{\gdelta}[0]{$\delta$}
\newcommand{\gepsilon}[0]{$\eps ilon$}
\newcommand{\gphi}[0]{$\phi$}
\newcommand{\gvarphi}[0]{$\varphi$}
\newcommand{\geta}[0]{$\eta$}
\newcommand{\gtheta}[0]{$\theta$}
\newcommand{\gomega}[0]{$\omega$}
\newcommand{\gchi}[0]{$\chi$}
\newcommand{\gxi}[0]{$\xi$}
\newcommand{\gkappa}[0]{$\kappa$}
\newcommand{\del}[1]{\partial_{#1}}
\newcommand{\av}[1]{\langle #1\rangle}
\newcommand{\bra}[1]{\langle #1\right|}
\newcommand{\ket}[1]{\left| #1\right\rangle}
\newcommand{\bracket}[3]{\langle #1 | #2 | #3\rangle}
\newcommand{\sproduct}[2]{\langle #1 | #2\rangle}
\newcommand{\Bracket}[3]{\left\langle #1 \left| #2 \right| #3\right\rangle}
\newcommand{\sign}[1]{\mbox{sign}\left(#1\right)}
\newcommand{\mc}[3]{\multicolumn{#1}{#2}{#3}}
\newcommand\T{\rule{0pt}{2.6ex}}
\newcommand\B{\rule[-1.2ex]{0pt}{0pt}}

\title{
\textit{Ab initio} structure modeling of complex thin-film oxides:
thermodynamical stability 
of TiC/thin-film alumina
}

\author{Jochen Rohrer}
\email{rohrer@chalmers.se}
\affiliation{%
BioNano Systems Laboratory, 
Department of Microtechnology, 
MC2, 
Chalmers University of Technology, 
SE-412 96 Gothenburg
}%
\author{Carlo Ruberto}%
\affiliation{%
BioNano Systems Laboratory, 
Department of Microtechnology, 
MC2, 
Chalmers University of Technology,
SE-412 96 Gothenburg
}%
\affiliation{%
Materials and Surface Theory Group, 
Department of Applied Physics, 
Chalmers University of Technology,
SE-412 96 Gothenburg
}%
\author{Per Hyldgaard}%
\affiliation{%
BioNano Systems Laboratory, 
Department of Microtechnology, 
MC2, 
Chalmers University of Technology,
SE-412 96 Gothenburg
}%

\date{\today}

\begin{abstract}
We present an efficient and general method  to identify promising candidate configurations for thin-film oxides 
and to determine structural characteristics of (metastable) thin-film structures  using \textit{ab initio} calculations. 
At the heart of this method is the complexity of the oxide bulk structure,
from which a large number of  thin films with structural 
building blocks, that is motifs,
from metastable bulk oxide systems can be extracted.
These span a coarse but well-defined network of initial configurations
for which density functional theory (DFT) calculations 
predict and implement dramatic atomic  relaxations 
in the corresponding, resulting thin-film candidates.
The network of thin-film candidates (for various film thicknesses and stoichiometries) 
can be ordered according to their variation in {\it ab initio\/} total energy 
or in \textit{ab initio\/} equilibrium Gibbs free energy.
Analysis of the relaxed atomic structures for the most favored structures
gives insight into the  nature of stable and metastable thin-film oxides. 
We investigate ultrathin alumina nucleated on TiC as a model system to illustrate this method.
The stable $\alpha$- and metastable \Kalumina\ bulk structures
lead to an alumina-film candidate-space that consists of $38$ configurations 
for a given film thickness, including three different stoichiometries.
We identify the stoichiometries that are relevant in equilibrium
with an O environment from \textit{ab initio} thermodynamics calculations 
of the relaxed configurations. 
These relevant stoichiometries are Al$_{4n-4}$O$_{6n}$ and 
Al$_{4n-2}$O$_{6n}$ (only in equilibrium at extremely low O chemical potentials),
with $n=2$, $3$, $4$ identifying the number of oxygen layers. 
The films with Al$_{4n}$O$_{6n}$ stoichiometry 
are not stable for any allowed value of the O chemical potential.
Our analysis of the atomic structure shows that the favorable structural
motifs of the relaxed films heavily differ from those in the bulk.
In particular the number of tetrahedrally coordinated Al ions is much higher in the films 
and the corresponding tetrahedra are oriented differently than in the bulk.
This finding of additional or novel favorable motifs documents
that the method is capable of catching thin-film candidates with a structural nature
that is not explicitly included in the network of initial thin-film configurations.
Our analysis also shows 
that the thermodynamically most stable TiC/Al$_{4n-4}$O$_{6n}$ systems 
decay into a partly decoupled  TiC/O/Al$_{4n-4}$O$_{6n-6}$ system,
with only a weak binding of the Al$_{4n-4}$O$_{6n-6}$ film on the TiC/O substrate.
\end{abstract}


\maketitle
\section{Introduction}
Understanding the atomic and 
electronic structure 
of thin-film oxides
is of significant industrial 
and fundamental importance
and a huge challenge at the same time.
Bulk oxides are characterized by a strong 
ionicity, which often results into a 
tendency for a  high structural flexibility
and an organization in a large number of different
stable and metastable 
phases.
Prominent examples can be found among aluminum oxides, \cite{ref:AlxOyStructures}
titanium oxides,\cite{ref:TixOyStructures}
vanadium oxides,\cite{ref:VxOyStructures}
or hafnium oxides.\cite{ref:HfxOyStructures}
For an ultra-thin film, the structural variety
of the oxide can be even 
larger.\cite{ref:Stierle_NiAl-Alumina,ref:Kresse_NiAl-Alumina} 
The mainly insulating character
of oxides makes accurate
experimental atomic and electronic structure
determinations difficult,
since   high-resolution techniques 
(low energy electron diffraction (LEED), \cite{ref:Stierle_NiAl-Alumina} 
scanning tunneling microscopy (STM),\cite{ref:Kresse_NiAl-Alumina}
transmission electron microscopy  (TEM), \cite{ref:TEM}
scanning electron microscopy(SEM),\cite{ref:SEM}  \ldots)
mainly use charged particles.
Theory assisted methods,
such as  density functional theory 
(DFT) calculations,
are of high complementary value.
However, when modeling 
thin films that are
adsorbed on a  substrate,
relatively large surface unit cells are often needed.
As a consequence, an enormous number of possible 
atomic configurations for the film 
arises,
and a structure determination by straightforward
energy calculations of all possible candidates 
becomes computationally intractable.

The nucleation of alumina on TiC provides an illustration of the complexity and 
importance of predicting and understanding atomic structure in  oxides,
ultra-thin oxide films and their interfaces.
Multilayers of TiC/alumina are highly relevant for industrial 
application as wear-resistant coatings on cemented-carbide cutting 
tools.\cite{ref:TiX-Al2O3_Coatings}  
They are commonly fabricated 
by chemical vapor deposition (CVD).
Typically, the \Aalumina\ 
(stable in the bulk) 
and \Kalumina\ (metastable in the bulk) phases 
are obtained
with relative orientations 
are $\alpha(0001)||\mbox{TiC}(111)$ 
and $\kappa\{001\}||\mbox{TiC}(111)$.\cite{ref:Al2O3TiX_TEMGrowth}
However, these ordered structures only arise when the alumina
possesses a considerable thickness.
The nucleation of alumina on TiC involves 
the formation of ultra-thin alumina films. 
Insight into the detailed, 
atomic configuration in the ultra-thin films is essential because 
their structure may strongly influence  the subsequent growth.
\cite{ref:Rohrer_CoarseGrained}
A complete search
through all possible atomic
thin-film configurations 
by total energy calculations
is, however, extremely difficult.\cite{ref:NoteOnComplexity}

Of course, \textit{ab initio} molecular dynamics (MD)
\cite{ref:MD_general}
is a powerful tool that can generally sample typical, 
and hence relevant, 
structural configurations (for a given film thickness and stoichiometry) 
by simply following the atomic dynamics at 
some elevated temperature. 
The use of density functional theory (DFT) 
to evaluate the forces 
on the constituent atoms 
makes this a highly accurate, but also 
costly, method. 
For strongly ionic materials like (thin-film) alumina 
drastic charge rearrangements
occur with the motion of every single atom. 
As this calls  for very small time steps, 
formidable \textit{ab initio} MD simulation times can be expected. 
Moreover, for a study of thin-film nucleation 
it is imperative to explore possible candidate structures
for a range of different thicknesses and stoichiometric compositions.
It is clear that some alternative, accelerated search method 
is desirable.

In this paper, we suggest an \textit{ab initio} method to search
for the structural elements of 
thermodynamically stable and metastable 
thin-film oxide 
configurations nucleated on metallic substrates. 
The \textit{ab initio} method avoids costly molecular-dynamics simulations 
but invokes atomic relaxations specified by underlying
DFT calculations of the atomic forces. 
The approach makes use of the complexity of
oxide bulk structures and their ionicity.
The method is illustrated for the  TiC/thin-film alumina
system that we investigate as an example.
We start with a coarse sampling of the
thin-film configuration space
by relaxing all possible films
that consist of partial alumina
bulk structures.
Due to their stability in the
bulk they can be assumed
as promising initial configurations
that contain structural elements, that is  motifs,
relevant for the actual stable and metastable 
thin-film structures.
To further trigger relaxations towards
other configurations that 
do not possess a partial 
bulk structure,
we slightly distort 
the unrelaxed candidates.
The relaxed 
atomic structures that are 
energetically favorable
and that differ essentially from
the partial bulk structures
can be used to design new
candidates with a similar 
structure.
The procedure could  then
be iterated  until self-consistency is reached.
For thin-film configurations
that differ in their stoichiometry,
the energy criterion
is replaced by the criterion
of lowest Gibbs free
energy.
The latter we calculate from
\textit{ab initio} thermodynamics. \cite{ref:Kaxiras_Thermodynamics,ref:Finnis_Thermodynamics,ref:Scheffler_RuO2}

The paper is organized as follows:
Section~\ref{sec:Background} summarizes the
properties of alumina 
and TiC that are relevant for TiC/thin-film alumina.
In Sec.~\ref{sec:Modeling}
we derive all TiC/thin-film alumina initial configurations
that are consistent with the bulk structure
of the respective materials.
The details concerned with the
computation of total and Gibbs free energies
are discussed in Sec.~\ref{sec:ComputationalMethod}.
In Sec.~\ref{sec:Energetics},
we present our results on the energetics and thermodynamical
stability of thin-film alumina.
An analysis of the atomic structure
of relaxed films is given in Sec.~\ref{sec:Structure}.
In Sec.~\ref{sec:Discussion}, 
we discuss our results
and Sec.~\ref{sec:Summary},
contains our conclusions.

\section{Materials Background\label{sec:Background}}
We first summarize the bulk and surface
properties of alumina and TiC
that are relevant for our thin-film modeling method.

\subsection{Stable and metastable \alumina\ bulk structures}
\begin{figure}
\begin{tabular}{c}
\epsfig{file=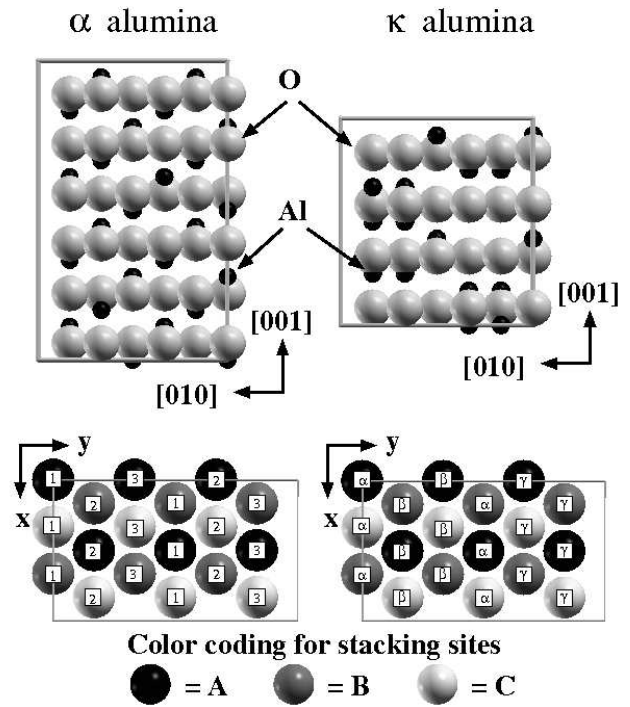,width=8.4cm}
\end{tabular}
\caption{Bulk structures 
of 
$\alpha$- (left) and \Kalumina\ (right) within 
orthorhombic unit cells.
The top panels show side views along 
$[100]$.
The bottom panels define 
the atomic site 
labeling within each $(001)$ atomic layer. 
}
\label{fig:Bulk}
\end{figure}

\begin{table}
\begin{ruledtabular}
\begin{tabular}{ccccccc}
Reflection about  & \mc{6}{c}{Effect on}\\
\B                & $A(a)$ & $B(b)$& $C(c)$& $\alpha(1)$&$\beta(2)$&$\gamma(3)$\\ 
\hline
\T
$xz$-plane& 
$A(a)$ & $B(b)$& $C(c)$& $\alpha(1)$&$\gamma(3)$&$\beta(2)$\\ 
$yz$-plane& 
$A(a)$ & $C(c)$& $B(b)$& $\alpha(1)$&$\beta(2)$&$\gamma(3)$
\end{tabular}
\end{ruledtabular}
\caption{\label{tab:Symmetries}
Mapping of stacking 
and site labels (as defined in Fig.\ \ref{fig:Bulk}) 
under mirror transformations.
A reflection about the 
$xz$-plane 
leads for example to a relabeling of 
$c_{\beta}\rightarrow c_{\gamma}$. 
}
\end{table}

Figure~\ref{fig:Bulk} shows a schematics of the bulk structures 
of  $\alpha$ (trigonal unit cell, space group $R\bar{3}c$)
and  $\kappa$ alumina (orthorhombic unit cell, space group 
$Pna2_1$).\cite{ref:alphaStructure,ref:Yashar_KappaBulk} 
Along the $\alpha[0001]$ and $\kappa[001]$ directions,
both alumina phases are composed
of alternating O and Al layers,
the latter splitting 
up 
into two 
sublayers. 
In \Aalumina, all Al ions are octahedrally ($O$)
coordinated.
In \Kalumina, 
the coordination alternates.
In every second layer all Al ions have 
octahedral coordination.
In the 
other 
layers 
$50$~\% of the Al ions
is octahedrally and $50$~\% tetrahedrally ($T$)
coordinated.
All tetrahedra point in the $[001]$ direction. 

To facilitate a parallel treatment of
$\alpha$- and \Kalumina,
we choose an orthorhombic unit cell 
for the representation of both alumina phases,
so that $\alpha[0001]_{\mtext{hex}}\Leftrightarrow
\alpha[001]_{\mtext{ortho}}$. 
The associated calculated lattice parameters are
$a=4.798$ ($4.875$)~\AA, $b=8.311$ ($8.378$)~\AA, 
and $c=13.149$ ($9.018$)~\AA\ for $\alpha$ 
($\kappa$),\cite{ref:Carlo_Surfaces,ref:Yashar_KappaBulk} 
which is in good agreement with experimental 
data.\cite{ref:Alpha_latticeParameterExp,ref:Kappa_latticeParameterExp} 

The atomic 
structures 
of 
$\alpha$- and \Kalumina\
can be described as follows.
We denote the stacking sites of full
O layers by capital letters.
For 
the 
Al layers small letters with subscript
(Arabic numerals for $\alpha$,
Greek letters for $\kappa$) are used.
The subscript relates each two Al sites per
unit cell, 
see 
Fig.~\ref{fig:Bulk} 
for a detailed definition of each label.
For $\kappa$ this notation is
identical to the one introduced in
Ref.~\onlinecite{ref:Yashar_KappaBulk},
whereas for $\alpha$
it is a slightly modified
version of the 
one of 
Ref.~\onlinecite{ref:Halvarsson_alpha},
where Greek superscripts are used for 
the labeling of Al vacancies.
With this notation, the bulk stackings 
are\cite{ref:Halvarsson_alpha,ref:Yashar_KappaBulk}
\bea
\alpha[0001]:&&~Ac_3c_2Bc_1c_3Ac_2c_1Bc_3c_2Ac_1c_3Bc_2c_1\nn\\
\kappa[001]:&&~Ab_{\gamma}c_{\beta}Bc_{\alpha}c_{\gamma}Ac_{\beta}b_{\gamma}Cb_{\alpha}b_{\beta}~.
\label{eq:BulkStacking}
\eea

Reflections about 
the $xz$-plane [$\Leftrightarrow(010)$] 
or $yz$-plane [$\Leftrightarrow (100)$]
are symmetries of the bulk.
Although the structure is not invariant
under these transformations,
the transformed structures
are equivalent to the 
non-transformed one.
The effects of the mirror transformations
on individual stacking sites
are listed in Table~\ref{tab:Symmetries}.
Reflection about the $xz$-plane
corresponds  to interchanging 
$\beta\leftrightarrow\gamma$
and $2\leftrightarrow3$, 
reflection about 
the $yz$-plane
to $C\leftrightarrow B$ and $c\leftrightarrow b$.

\subsection{TiC$(111)$ surface and reactivity \label{sec:TiCsurface}}
Bulk TiC possesses NaCl structure with a 
theoretical lattice parameter \cite{ref:AdsorptionOnTiX} 
$a=4.332$~\AA~
(in good agreement with the experimental 
value \cite{ref:TiCLatticeExp} $a_{\mtext{exp}}=4.33$~\AA).
Along the $[111]$ direction,
it is thus composed of close-packed alternating Ti and C layers.
The stacking sequence of one repeat unit is $ABCABC$.

We only consider Ti-terminated TiC(111) surfaces.
This choice is motivated by the stronger binding
of  Ti to the C-terminated surface 
compared to the binding of C to the 
Ti-terminated surface.\cite{ref:AdsorptionOnTiX} 
Furthermore, there is experimental evidence
for a preferred Ti termination upon 
annealing.\cite{ref:TiC_TerminationExp}

On  Ti-terminated TiC(111),
atomic O adsorbs much more strongly
than atomic Al (about three times as 
strong).\cite{ref:AdsorptionOnTiX} 
We therefore identify the first alumina layer 
above the TiC/\alumina\ interface plane 
as an O layer.
According to 
Refs.\ \onlinecite{ref:AdsorptionOnTiX} and \onlinecite{ref:Sead}, 
both single O atoms and a full O monolayer
prefer adsorption in the fcc site.
By defining the TiC stacking such that
the fcc site on its $(111)$ surface is labeled by 
an $A$ stacking letter, 
the hcp site by $B$, and the top site by $C$,
the position of the first O layer is therefore fixed to 
$A$ stacking. 
For the monolayer, our calculated 
Ti--O layer separation along TiC$[111]$ is 
$d_{\mtext{Ti-O}}=0.89$~\AA.


\begin{table*}
\begin{ruledtabular}
\begin{tabular}{lll}
\underline{\mbox{TiC/$\alpha[0001]$}}&
\underline{\mbox{TiC/$\kappa[001]$}}&
\underline{\mbox{TiC/$\kappa[00\bar{1}]$}}\\
\mbox{TiC/$Ac_3c_2Bc_1c_3Ac_2c_1Bc_3c_2Ac_1c_3Bc_2c_1$}&
\mbox{TiC/$Ab_{\gamma}c_{\beta}Bc_{\alpha}c_{\gamma}Ac_{\beta}b_{\gamma}Cb_{\alpha}b_{\beta}$}&
\mbox{TiC/$Ab_{\beta}b_{\alpha}Cb_{\gamma}c_{\beta}Ac_{\gamma}c_{\alpha}Bc_{\beta}b_{\gamma}$}\\
\mbox{TiC/$Ab_2b_3Cb_1b_2Ab_3b_1Cb_2b_3Ab_1b_2Cb_3b_1$}&
\mbox{TiC/$Ab_{\alpha}b_{\gamma}Cb_{\beta}a_{\gamma}Ba_{\alpha}a_{\beta}Ca_{\gamma}b_{\beta}$}&
\mbox{TiC/$Ab_{\beta}a_{\gamma}Ca_{\beta}a_{\alpha}Ba_{\gamma}b_{\beta}Cb_{\gamma}b_{\alpha}$}\\
&
\mbox{TiC/$Ac_{\beta}b_{\gamma}Cb_{\alpha}b_{\beta}Ab_{\gamma}c_{\beta}Bc_{\alpha}c_{\gamma}$}&
\mbox{TiC/$Ac_{\gamma}c_{\alpha}Bc_{\beta}b_{\gamma}Ab_{\beta}b_{\alpha}Cb_{\gamma}c_{\beta}$}\\
&
\mbox{TiC/$Ac_{\alpha}c_{\beta}Bc_{\gamma}a_{\beta}Ca_{\alpha}a_{\gamma}Ba_{\beta}c_{\gamma}$}&
\mbox{TiC/$Ac_{\gamma}a_{\beta}Ba_{\gamma}a_{\alpha}Ca_{\beta}c_{\gamma}Bc_{\beta}c_{\alpha}$}
\end{tabular}
\end{ruledtabular}
\caption{\label{tab:Interfaces}
TiC(111)/alumina interface configurations that 
respect the bulk structure of $\alpha$- and \Kalumina\
and start with an O layer in fcc 
site 
on the 
Ti-terminated
TiC(111). 
The TiC stacking of 
the surface region 
is defined as $\ldots ABCABC$.
}
\end{table*}

\section{Thin-Film Identification Method\label{sec:Modeling}}
We seek a characterization of alumina nucleation on TiC and 
identify promising thin-film alumina structure-candidates 
for different oxide layer thicknesses and stoichiometries.
This section describes in detail the proposed method
for finding promising initial configurations for TiC/thin-film alumina
candidates. 
First, we derive all 
the TiC/alumina interface configurations
that are consistent with the
respective bulk structures
and that take into account the
adsorption properties
of TiC$(111)$.
Then, we obtain all the initial thin-film 
configurations that consist of partial bulk alumina
by truncating these interface sequences.

\subsection{TiC/alumina interface structures\label{sec:Interfaces}}
Table~\ref{tab:Interfaces}
lists all 
the TiC/\Aalumina\ and TiC/\Kalumina\  
interfaces that are conform with
the respective bulk structures
and that start 
with an O layer
in fcc ($A$) 
site on the 
Ti-terminated 
TiC $(111)$ 
surface.
These stacking sequences are found as follows:

We observe that any of the O layers in the listing~(\ref{eq:BulkStacking})
can be chosen as the initial alumina layer. 
This layer must be translated to
an $A$ site, which can be achieved
by  cyclic permutations. 
For the $C$ sites to be translated to $A$ sites we 
need one cyclic permutation,
for $B$ sites to be translated to $A$ sites we 
need two.  
All other sites are relabeled accordingly.
For example, for $C\rightarrow A$, we have
$A\rightarrow B$ and $B\rightarrow C$.  
For the Al positions the 
corresponding relabeling has to be performed, 
keeping the subscripts [$\alpha$ ($1$), \ldots] 
fixed.

Next, we note that the fixed stacking sequence 
of the TiC substrate breaks the symmetry
associated with a reflection about 
the $yz$-plane.
Hence, for each alumina sequence,
we need to consider an additional one,
which is obtained by interchanging
$B(b) \leftrightarrow C(c)$.

Finally, we exploit that reflection about 
the $xz$-plane is still a symmetry of TiC/alumina
since the TiC is composed of fully occupied
layers.
Hence, alumina sequences that
are related by $\beta\leftrightarrow\gamma$
($2\leftrightarrow3$)
are equivalent.

For \Aalumina, all O layers are 
equivalent.
Therefore it is sufficient to focus on the first 
O layer, which is already in 
$A$ stacking. 
Also, $\alpha[001]\Leftrightarrow\alpha[00\bar{1}]$ 
and thus 
only 
the symmetry breaking associated with 
the reflection about the $yz$-plane needs 
to be considered.  As a result, only two possible 
interfacial configurations have to be taken into 
account (see Table~\ref{tab:Interfaces}, left column). 

For \Kalumina, only every second O layer is equivalent 
and $\kappa[001]$ is not equivalent to 
$\kappa[00\bar{1}]$.  
Therefore, we need 
to consider both directions,
any two consecutive bulk O layers,
and the effect of the symmetry breaking. 
This results in four different configurations 
for each direction (see Table ~\ref{tab:Interfaces}, middle 
and right columns).

\subsection{TiC/thin-film alumina candidate structures}
We obtain the network of initial
thin-film alumina configurations
in three steps.
In the first step, we truncate the TiC/alumina interface sequences
in Table~\ref{tab:Interfaces} after a full Al layer.
The number of O layers $n$ defines the thickness of the film. 
In a second step, the resulting configurations are distorted 
by placing the Al sublayers 
into one and the same plane, 
exactly in between 
the two neighboring O layers.
In the third step, 
we vary the stoichiometry by removing Al ions from the 
surface in accordance with the bulk space group,
\textit{i.e.}, 
only  Al pairs that belong to the same stacking label are removed.

In this way, for each thickness
we generate three stoichiometry classes:
Al$_{4n}$O$_{6n}$, Al$_{4n-2}$O$_{6n}$, and Al$_{4n-4}$O$_{6n}$, 
corresponding to the removal of zero, one, and two Al pairs, 
respectively.  

For the $\alpha$-Al$_{4n-2}$O$_{6n}$ films, 
we only consider the
surface Al pair that has no direct neighbor 
in the layer below.
With this choice, we minimize the
electrostatic energy.
We have confirmed the quality of
this choice by comparing
the total energy for some configurations
with different choices of the surface Al pair.

For the $\kappa$-Al$_{4n-2}$O$_{6n}$ films,
there is no such simple argument
and we choose to allow for both
possible Al pairs.
As a result, the number of 
$\kappa$-Al$_{4n-2}$O$_{6n}$ 
configurations
is twice as large as the number of 
$\kappa$-Al$_{4n-4}$O$_{6n}$ 
or $\kappa$-Al$_{4n}$O$_{6n}$
configurations.

In Tables~\ref{tab:Erel_O-class}--\ref{tab:Erel_AlAl-class},
all candidates found by the described
procedure are listed for 
films with thicknesses $n = 2$, $3$, and $4$ 
(for thermodynamical reasons,
see Sec.~\ref{sec:Thermodynamics}, no 
Al$_{16}$O$_{24}$ 
configurations are considered).


\section{\textit{Ab initio}  Method\label{sec:ComputationalMethod}}

\subsection{Total energies and atomic relaxations \label{sec:CompMethod}}
All calculations are performed with the DFT plane-wave code dacapo \cite{ref:Dacapo}
using ultra-soft 
pseudopotentials 
\cite{ref:PSP} 
and the PW91 exchange-correlation \cite{ref:PW91}
functional.

We use a supercell approach 
and model the TiC/thin-film alumina by slab
geometry. 
The basal plane dimensions 
of the supercell 
are chosen to fit the $3\times2$ TiC(111) surface
($5.306\times9.190$~\AA$^2$)
and the height is fixed to 
$30$~\AA , 
ensuring a vacuum thickness of at 
least $13$~\AA . 

The TiC is modeled by 
four atomic 
bilayers (with six Ti and 
six C atoms per bilayer).  
The alumina films contain 
six O atoms per O layer
and a varying number of Al atoms, 
depending on the film stoichiometry. 
In total, the slabs contain between $64$ 
(Al$_{4}$O$_{12}$ films)
and $86$ atoms 
(Al$_{14}$O$_{24}$ films).

We use a 400~eV plane-wave cutoff and a 4$\times$2$\times$1 Monkhorst-Pack\cite{ref:MP}
$k$-point sampling. 
Electrostatic effects arising from the charge asymmetry in the slab
are corrected for by a dipole correction.
The atomic relaxations are performed 
until all interatomic forces are smaller than 0.05~eV/\AA.
This choice has proven a good accuracy at acceptable CPU times for
$\alpha$- and \Kalumina\ 
surfaces\cite{ref:Carlo_Surfaces} 
and for TiC/alumina interface calculations.~\cite{ref:Carlo_PhD}
The presented DFT calculations amount to a total of 
one million CPU hours on modern supercomputing facilities.

\subsection{Equilibrium Thermodynamics}
\begin{figure}
\begin{tabular}{c}
\epsfig{file=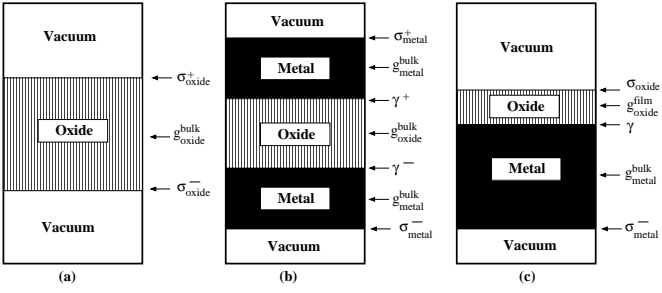,width=8.8cm}
\end{tabular}
\caption{\label{fig:Modeling}
Schematics of atomic setup for calculations of 
(a)~(oxide) surfaces (b)~(metal/oxide) interfaces 
and (c)~thin-film oxide on a metal substrate. 
The arrows point to the regions in the
slabs where the bulk ($g^{\mtext{bulk}}$), surface ($\sigma$), interface ($\gamma$)
or thin-film ($g^{\mtext{film}}$) contributions 
to the Gibbs free energy are located.
We emphasize that, in general 
$g^{\mtext{bulk}}_{\mtext{oxide}}\neq g^{\mtext{film}}_{\mtext{oxide}}$.
}
\end{figure}

At non-zero temperature $T$ and pressure $p$ 
the stability of any system 
is governed by the Gibbs free energy  $G$.

\textit{Surfaces. } 
Figure ~\ref{fig:Modeling}(a) shows a typical atomic setup used
for calculating surface energies using slab geometry. \cite{ref:alphaSurfaces_Thermodynamics}
For alumina, the average surface Gibbs free energy $\sigma_{\mtext{av}}$ 
of the pair of  alumina surfaces represented by the slab
is defined by 
\bea
\sigma_{\mtext{av}}=\frac{1}{2}\left(\sigma^++\sigma^-\right)=
\frac{1}{2A}
\left(G_{\mtext{slab}}-
n_{\mtext{Al}}\mu_{\mtext{Al}}-n_{\mtext{O}}\mu_{\mtext{O}}\right)~.
\label{eq:GibbsSurfaceInit}
\eea
Here $\sigma^+$ and $\sigma^-$ correspond to the two surface energies
associated with each side of the slab
($\sigma_{\mtext{av}}=\sigma^+=\sigma^-=\sigma$ for a symmetric slab),
$G_{\mtext{slab}}$ is the Gibbs free energy of the
slab that contains $n_{\mtext{Al}}$ Al
and $n_{\mtext{O}}$ O atoms,
and $\mu_{\mtext{Al}}$ and $\mu_{\mtext{O}}$
are the chemical potentials of Al and O respectively.

In equilibrium with an O environment,
the stoichiometrically weighted sum of the
Al and O chemical potentials must equal the
Gibbs free energy per stoichiometric unit of alumina
$g_{\mtext{\alumina}}$,
\bea 
2\mu_{\mtext{Al}}+3\mu_{\mtext{O}}=
g_{\mtext{\alumina}}~,
\label{eq:EquilibriumStandard}
\eea
so that $\sigma_{\mtext{av}}$ can be rewritten as a function
of the O chemical potential only.

\textit{Interfaces. }
Figure ~\ref{fig:Modeling}(b) shows a schematics of 
of a typical slab geometry used to calculate 
interface energies. \cite{ref:Zhang_AluminaInterface}
For metal/alumina interfaces, 
the stability is determined by the interface Gibbs free energy $\gamma$.
Using the slab geometry of Fig.~\ref{fig:Modeling}(b),
the average interface energy is calculated as
\bea
\gamma_{\mtext{av}}&=&
\frac{1}{2}\left(\gamma^++\gamma^-\right) \nn \\ 
&=&\frac{1}{2A} 
\left(G_{\mtext{slab}}-\sum_in_i\mu_i 
-A\sigma_{\rm metal}^+-A\sigma_{\rm metal}^-\right)\nn\\
&=&\frac{1}{2A}
\bigg(G_{\mtext{slab}}-n_{\mtext{metal}}g_{\mtext{metal}}-
\frac{n_{\mtext{Al}}}{2}g_{\mtext{\alumina}}-\nn\\
&&\left(n_{\mtext{O}}-\frac{3n_{\mtext{Al}}}{2}\right)\mu_{\mtext{O}}
-A\sigma_{\rm metal}^+-A\sigma_{\rm metal}^-
\bigg)~.
\label{eq:GibbsInterface}
\eea
Here $\gamma^+$ and $\gamma^-$
are the interface energies corresponding to
the two interfaces in Fig.~\ref{fig:Modeling}(a),
$G_{\mtext{slab}}$ is the Gibbs free energy of the total slab,
and $n_i$ and  $\mu_i$ are the number and chemical potentials 
of the different atomic species ($i\in\{\text{Al, O, or metal}\}$).
In the last equality, we have assumed equilibrium with an O environment
and used Eq.~\ref{eq:EquilibriumStandard} to express 
$\gamma_{\mtext{av}}$ as a function of the O chemical potential only.
For a symmetric slab, we have $\gamma_{\mtext{av}}=\gamma^+=\gamma^-=\gamma$.

\textit{TiC/thin-film alumina. }
Figure \ref{fig:Modeling}(c) sketches 
the atomic setup used for the present thin-film calculations.
The stability of TiC/thin-film alumina is governed 
by three contributions:
(i)~the surface stability of the thin film ($\sigma$), 
(ii)~the stability of the interfacial configuration ($\gamma$),
and (iii)~the internal stability of the thin film itself ($g^{\mtext{film}}_{\mtext{\alumina}}$).

The first two contributions can be expressed by 
\bea
A(\gamma+\sigma)&=&
G_{\mtext{slab}}-n_{\mtext{TiC}}g_{\mtext{TiC}}-
n_{\mtext{Al}}\mu_{\mtext{Al}}-
n_{\mtext{O}}\mu_{\mtext{O}}-\nn\\
&&A\sigma_{\rm TiC}^-~.
\label{eq:GibbsThinFilm}
\eea
Because we always consider the same TiC slab (always having identical surface energies)
it is convenient to express $n_{\mtext{TiC}}g_{\mtext{TiC}}$ through the calculated
Gibbs free energy of a TiC slab
\bea
G_{\mtext{TiC}}=n_{\mtext{TiC}}g_{\mtext{TiC}}+A\sigma_{\rm TiC}^++A\sigma_{\rm TiC}^-~.
\eea
The stability of the thin-film system is therefore conveniently described by 
the Gibbs free energy difference
\bea
\Gamma&=&A(\gamma+\sigma-\sigma^-_{\mtext{TiC}}-2\sigma^+_{\mtext{TiC}})= \nn\\
&=&
G_{\mtext{TiC/alumina}}-G_{\mtext{TiC}}-
n_{\mtext{Al}}\mu_{\mtext{Al}}-
n_{\mtext{O}}\mu_{\mtext{O}} 
\label{eq:GibbsPrel}
\eea
Note that 
for practical reasons we do not normalize
$\Gamma$ by the  area of the basal plane
of the unit cell, 
since our unit cells all have the same 
basal plane.

For the third contribution that governs
the stability of thin films,
$g^{\mtext{film}}_{\mtext{\alumina}}$,
we note that in general the Gibbs free energy per stoichiometric unit
may and is expected to differ
from that in the bulk,
that is,
$g^{\mtext{film}}_{\mtext{\alumina}}\not= g_{\mtext{Al$_2$O$_3$}}$.
We therefore introduce a parameter $\delta$,
which measures the difference between
the Gibbs free energy of one stoichiometric unit 
of alumina in a bulk environment and in the film
and rewrite the equilibrium condition as
\bea
2\mu_{\mtext{Al}}+3\mu_{\mtext{O}}=g_{\mtext{\alumina}}+\delta~.
\label{eq:equilibrium}
\eea
The stability-determining quantity for
TiC/thin-film alumina configurations 
can be reformulated from Eqs.~\ref{eq:GibbsPrel} and \ref{eq:equilibrium}:
\bea
\Gamma&=&G_{\mtext{TiC/Al$_n$O$_m$}}
-G_{\mtext{TiC}}
-\frac{n_{\mtext{Al}}}{2}\left(g_{\mtext{\alumina}}+\delta\right)\nn\\
&&-(n_{\mtext{O}}-\frac{3}{2}~n_{\mtext{Al}})~\mu_{\mtext{O}}~.
\label{Eq:Gibbs}
\eea
The limits of the physically
allowed range of the chemical potentials
are defined by Al condensation into fcc Al
and  O condensation  into O$_2$,
\textit{i.e.}
$\mu_{\mtext{Al}}<g_{\mtext{fcc-Al}}$
and
$\mu_{\mtext{O}}<\frac{1}{2}\mu_{\mtext{O$_2$}}$,
where $g_{\mtext{fcc-Al}}$
and $\mu_{\mtext{O$_2$}}$
are the Gibbs free energy per stoichiometric unit of fcc Al
and the chemical potential of  
O$_2$, respectively.
Combining both inequalities 
and Eq.~\ref{eq:equilibrium} 
yields
\bea 
\frac{1}{3} (g_{\mtext{\alumina}}+\delta-2g_{\mtext{fcc-Al}})< 
\mu_{\mtext{O}}<\frac{1}{2}\mu_{\mtext{O$_2$}}.
\label{Eq:mu-range} 
\eea

Although an exact value of $\delta$ cannot be calculated,
we can estimate $\delta$ by calculating
energy differences between films that differ
by an integer number of stoichiometric units,
\bea
\delta_{nm}=
(E_{\mtext{Al}_n\mtext{O}_m}-E_{\mtext{Al}_{n-4}\mtext{O}_{m-6}}-2\epsilon_{\mtext{\alumina}})/2~,
\eea
We find that $\delta_{nm}$ 
is  $0.4$~eV and $0.7$~eV 
when comparing three and two
and four and three layer thick films
for Al$_{4n-4}$O$_{6n}$ stoichiometry.
For Al$_{4n-2}$O$_{6n}$ stoichiometry
the corresponding values are
$0.3$~eV and $1.2$~eV.
For Al$_{4n}$O$_{6n}$ stoichiometry
we have only considered 
three and two layer thick films for which
we find $\delta_{nm}=1.1$~eV.

The fact that the largest values of $\delta_{nm}$ 
are found when calculating the energy differences
for the thickest considered films
is counterintuitive.
We would expect that the difference in Gibbs free energy
per stoichiometric units converges towards that of the bulk
once the film is thick enough.
This shows the difficulties in determining
the Gibbs free energy of a thin film properly.
The higher values for thicker films
may be due to completely different surfaces
of the respective films
and thus due to surface energies.

In the following, we disregard the fact of a non-zero value
and the stoichiometry and thickness  dependence of $\delta$,
that is, we put $\delta\equiv0$.
We have checked that the resulting
uncertainty in $\Gamma$, 
although certainly not negligible,
does not change our qualitative results
as long as the temperatures are not too high 
(below $1300$~K).

\subsection{\textit{Ab initio}  Equilibrium Thermodynamics}
For the calculation of the Gibbs free energies
and chemical potentials involved in Eq.~\ref{Eq:Gibbs}
we use the method described in 
Ref.~\onlinecite{ref:Scheffler_RuO2}.
Here, we only summarize their arguments and give the
computational prescription.

The Gibbs free energy of solid material,
that is of the \textit{bulk} or a \text{slab},
is essentially independent of the pressure.
The temperature dependence is considerably stronger
and larger in absolute value.
Based on the calculated vibrational surface Gibbs free energy
for RuO$_2$, \cite{ref:Scheffler_RuO2}  
we estimate the vibrational Gibbs free energy per cell
for alumina at $T=1000$~K to $\Gamma^{\mtext{vib}}\sim1-2$~eV.
Compared to the uncertainty in $\Gamma$ due the uncertainty
in $\delta$, the vibrational contributions
can therefore savely be neglected.\cite{ref:GammaVib}

We consequently choose to approximate $\Gamma$ by 
\bea
\Gamma&\equiv& 
E_{\mtext{TiC/Al$_n$O$_m$}}-E_{\mtext{TiC}}
-\frac{n_{\mtext{Al}}}{2}\epsilon_{\mtext{\alumina}}-\nn\\
&&(n_{\mtext{O}}-\frac{3}{2}~n_{\mtext{Al}})~\mu_{\mtext{O}}~,
\label{Eq:Gibbs_0}
\eea
where $E_{\mtext{TiC/Al$_n$O$_m$}}$, $E_{\mtext{TiC}}$, and $\epsilon_{\mtext{\alumina}}$
are DFT total energies of a TiC/Al$_n$O$_m$ slab,
an isolated (clean) TiC slab and one stoichiometric unit of bulk alumina respectively.

To estimate the partial O$_2$ pressure and the temperature
that correspond to different values of $\mu_{\mtext{O}}$,
we adopt the ideal gas approximation.
This approximation allows us to rewrite $\mu_{\mtext{O}}$
as 
\bea
\mu_{\mtext{O}}(T,p)&=&
\frac{1}{2}\left[\epsilon^{\mtext{DFT}}_{\mtext{O}_2}+\delta\mu_{\mtext{O}_2}(T,p_0)
+k_{\mtext{B}}T\ln\frac{p}{p_0}\right].
\label{Eq:mu_O} 
\eea 
Here $\epsilon^{\mtext{DFT}}_{\mtext{O}_2}$ is the DFT total energy
of the O$_2$ molecule and
$\delta\mu(T,p_0)$ is related to the
entropy $S$ and enthalpy $H$ at 
a fixed pressure $p_0$,
see Ref.~\onlinecite{ref:Scheffler_RuO2} for details.
To calculate $\delta\mu(T,p_0)$, we use the
values of $S$ and $H$  
for different temperatures 
at standard pressure $p_0=1$~atm
that are tabulated in  Ref.~\onlinecite{ref:JANAF}.

\begin{table*}
\begin{ruledtabular}
\begin{scriptsize}
\begin{tabular}{l|llr|llr|llr}
&\mc{3}{c|}{\underline{
Al$_4$O$_{12}$ 
films}}&
\mc{3}{c|}{\underline{
Al$_8$O$_{18}$ 
films}}&
\mc{3}{c}{\underline{
Al$_{12}$O$_{24}$ 
films}}\\
alumina& 
alumina& coord. of & $E_{\mtext{rel}}$&
alumina& coord. of & $E_{\mtext{rel}}$&
alumina& coord. of & $E_{\mtext{rel}}$\\
\B group  &
stacking& Al ions& (eV/cell)&
stacking& Al ions& (eV/cell)&
stacking& Al ions& (eV/cell)\\
\hline
\T$\alpha$& 
 $Ac_3c_2B$ & $OO$ &0.71 &
{$ Ac_3c_2Bc_1c_3A$}& $OO:OO$ &{5.02} &   
{$ Ac_3c_2Bc_1c_3Ac_2c_1B $}& $OO:OO:OO$  &   {2.34}    \\
\B$\alpha$& 
$ Ab_2b_3C$& $OO$ &0.72 &
{$ Ab_2b_3Cb_1b_2A $}& $OO:OO$&{2.78}&
{$ Ab_2b_1Cb_1b_2Ab_3b_1C $} & $OO:OO:OO$ & {2.19}\\  
\hline
\T$\kappa [001]$& 
$Ab_{\gamma}c_{\beta}B$ & $T_{\uparrow}O$ &{8.02} &
{$ Ab_{\gamma}c_{\beta}Bc_{\alpha}c_{\gamma}A $}& $T_{\uparrow}O:OO$ &{9.03}&
{$ Ab_{\gamma}c_{\beta}Bc_{\alpha}c_{\gamma}Ac_{\beta}b_{\gamma}C$ }&$T_{\uparrow}O:OO:T_{\uparrow}O$ &{7.15}  \\ 
$\kappa [001]$&
\underline{$Ab_{\gamma}c_{\beta}C$}&$T_{\uparrow}O$ &{\underline{0.01}}&
{$ Ac_{\beta}b_{\gamma}Cb_{\alpha}b_{\beta}A $}&$T_{\uparrow}O:OO$ &{2.46}  &
{$ Ac_{\beta}b_{\gamma}Cb_{\alpha}b_{\beta}Ab_{\gamma}c_{\beta}B$}&$T_{\uparrow}O:OO:T_{\uparrow}O$ &{7.26} \\
[0.25cm] 
$\kappa [001]$& 
$Ac_{\alpha}c_{\beta}B$  & $OO$ &{4.43} &
{$ Ac_{\alpha}c_{\beta}Bc_{\gamma}a_{\beta}C$} &  $OO:T_{\uparrow}O$&{2.03} &
{$ Ac_{\alpha}c_{\beta}Bc_{\gamma}a_{\beta}Ca_{\alpha}a_{\gamma}B $}&$OO:T_{\uparrow}O:OO$&{4.05}\\
$\kappa [001]$& 
$Ab_{\alpha}b_{\gamma}C$  & $OO$ &{2.43} &
{$ Ab_{\alpha}b_{\gamma}Cb_{\beta}a_{\gamma}B$}& $OO:T_{\uparrow}O$ &{1.39} &
{\B$ Ab_{\alpha}b_{\gamma}Cb_{\beta}a_{\gamma}Ba_{\alpha}a_{\beta}C $}&$OO:T_{\uparrow}O:OO$& {3.55} \\
\hline
\T$\kappa [00\bar{1}]$& 
\underline{$Ac_{\gamma}a_{\beta}B$}  & $OT_{\downarrow}$ &{\underline{0.20}}&
{$ Ac_{\gamma}a_{\beta}Ba_{\gamma}a_{\alpha}C   $ } &$OT_{\downarrow}:OO$ &{2.69} &
{$ Ac_{\gamma}a_{\beta}Ba_{\gamma}a_{\alpha}Ca_{\beta}c_{\gamma}B $ }  & $OT_{\downarrow}:OO:OT_{\downarrow}$   & {5.37}\\
$\kappa [00\bar{1}]$& 
\underline{$Ab_{\beta}a_{\gamma}C$}  & $OT_{\downarrow}$ &{\underline{0.00}} &
\underline{$ Ab_{\beta}a_{\gamma}Ca_{\beta}a_{\alpha}B$}  &$OT_{\downarrow}:OO$&{\underline{0.00}}&
{$ Ab_{\beta}a_{\gamma}Ca_{\beta}a_{\alpha}Ba_{\gamma}b_{\beta}C $}&$OT_{\downarrow}O:OO:OT_{\downarrow}$ &{4.15}\\
[0.25cm] 
$\kappa [00\bar{1}]$& 
$Ac_{\gamma}c_{\alpha}B$ & $OO$  &{4.43} &
{$ Ac_{\gamma}c_{\alpha}Bc_{\beta}b_{\gamma}A$ }& $OO:OT_{\downarrow}$&{2.79}  &
\underline{$ Ac_{\gamma}c_{\alpha}Bc_{\beta}b_{\gamma}Ab_{\beta}b_{\alpha}C$}     & $OO:OT_{\downarrow}:OO$  & {\underline{0.12}} \\
$\kappa [00\bar{1}]$& 
$Ab_{\beta}b_{\alpha}C$  & $OO$ &{2.43} &
{$ Ab_{\beta}b_{\alpha}Cb_{\gamma}c_{\beta}A  $ }&$OO:OT_{\downarrow}$ &{1.48}  &
\B\underline{$ Ab_{\beta}b_{\alpha}Cb_{\gamma}c_{\beta}Ac_{\gamma}c_{\alpha}B$ }&$OO:OT_{\downarrow}:OO$ & {\underline{0.00}}
\end{tabular}
\end{scriptsize}
\end{ruledtabular}
\caption{\label{tab:Erel_O-class}
Stacking sequence and Al coordination 
[$O$ for octahedral, $T$ for
tetrahedral, with the arrow indicating
the direction in which 
each tetrahedron vertex is pointing: 
towards the film surface ($\uparrow$) 
or towards the TiC/film interface ($\downarrow$)] 
of \textit{unrelaxed}  alumina films 
with Al$_{4n-4}$O$_{6n}$ stoichiometry 
and their total energies $E_{\mtext{rel}}$
\textit{after} relaxation
(given relative to the structure with lowest 
total energy for each film thickness). 
The configurations are grouped together according to 
the phase and orientation of the alumina bulk structures 
from which they are derived (left column). 
Configurations that differ only by a rotation of 
$180^{\circ}$ around TiC$[111]$ are organized into subgroups
separated by larger whitespace. 
In general, the 
unrelaxed and relaxed atomic 
structures 
differ considerably. 
The stable 
and potentially metastable 
(see text for details) 
configurations are underlined.
The \textit{ab initio} study and comparison permit us to make the
following set of observations:
(i)~The unrelaxed 
configurations with an
$AC$ stacking in the first two O layers
yield relaxed structures that are 
in general more favorable than
those obtained from configurations 
in which 
the stacking sequence 
has been rotated by 
$180^{\circ}$ around TiC$[111]$ 
($AB$ O stacking); 
(ii)~While for 
the Al$_4$O$_{12}$ 
films 
two different 
unrelaxed structures 
lead to
the stable configuration, 
for the other two film thicknesses 
only one 
structure leads to the stable configuration; 
(iii)~In general, the stable configurations are obtained from 
TiC$[111]$/$\kappa[00\bar{1}]$ initial structures; 
(iv)~The 
$\alpha$-type films 
lead to neither stable nor metastable configurations; and 
(v)~While 
the stable 
Al$_4$O$_{12}$ and Al$_8$O$_{18}$ films 
are 
both 
obtained from the same unrelaxed interface sequence (same line),
the stable 
Al$_{12}$O$_{24}$ film 
derives from another interface sequence.
}
\end{table*}

\begin{table*}
\begin{ruledtabular}
\begin{scriptsize}
\begin{tabular}{l|llr|llr|llr}
&\mc{3}{c|}{\underline{
Al$_6$O$_{12}$ 
films}}&
\mc{3}{c|}{\underline{
Al$_{10}$O$_{18}$ 
films}}&
\mc{3}{c}{\underline{
Al$_{14}$O$_{24}$ 
films}}\\
alumina& 
alumina& coord. of & $E_{\mtext{rel}}$&
alumina& coord. of & $E_{\mtext{rel}}$&
alumina& coord. of & $E_{\mtext{rel}}$\\
group  &
stacking& Al ions& (eV/cell)&
stacking& Al ions& (eV/cell)&
stacking& Al ions& (eV/cell)\\
\hline
\T$\alpha$& 
$ Ac_3c_2Bc_1 $&$OO:O$ &0.78 &
\underline{$ Ac_3c_2Bc_1c_3Ac_2$}&$OO:OO:O$ &\underline{0.15}  &
$  Ac_3c_2Bc_1c_3Ac_2c_1Bc_3$&   $OO:OO:OO:O$    & 0.99\\
$\alpha$& 
\B\underline{$Ab_2b_3Cb_1$} &$OO:O$ &\underline{0.02} &
$Ab_2b_3Cb_1b_2Ab_3$ & $OO:OO:O$ &3.22&
$Ab_2b_3Cb_1b_2Ab_3b_1Cb_2$& $OO:OO:OO:O$  & 0.97 \\
\hline
{\T$\kappa [001]$}& 
{$ Ab_{\gamma}c_{\beta}Bc_{\alpha} $}& $T_{\uparrow}O:O$ &{0.77} &
{$ Ab_{\gamma}c_{\beta}Bc_{\alpha}c_{\gamma}Ab_{\gamma}$}&$T_{\uparrow}O:OO:O$&{1.21 }&
{$ Ab_{\gamma}c_{\beta}Bc_{\alpha}c_{\gamma}Ac_{\beta}b_{\gamma}Cb_{\alpha} $}&$T_{\uparrow}O:OO:T_{\uparrow}O:O$&{2.89}\\
{$\kappa [001]$}& 
{$ Ab_{\gamma}c_{\beta}Bc_{\gamma}   $}&$T_{\uparrow}O:O$   &{4.38} &
\underline{$ Ab_{\gamma}c_{\beta}Bc_{\alpha}c_{\gamma}Ac_{\beta}$}&$T_{\uparrow}O:OO:T_{\uparrow}$&{\underline{0.00}}&
{$ Ab_{\gamma}c_{\beta}Bc_{\alpha}c_{\gamma}Ac_{\beta}b_{\gamma}Cb_{\beta}$}&$T_{\uparrow}O:OO:T_{\uparrow}O:O$&{2.89}\\
[0.25cm] 
{$\kappa [001]$}& 
\underline{$ Ac_{\beta}b_{\gamma}Cb_{\alpha}  $}&  $T_{\uparrow}O:O$&{\underline{0.14}} &
{$ Ac_{\beta}b_{\gamma}Cb_{\alpha}b_{\beta}Ac_{\beta}$}&$T_{\uparrow}O:OO:O$&{3.22}& 
{$ Ac_{\beta}b_{\gamma}Cb_{\alpha}b_{\beta}Ab_{\gamma}c_{\beta}Bc_{\alpha}$} &$T_{\uparrow}O:OO:T_{\uparrow}O:O$&2.63\\
{$\kappa [001]$}& 
\underline{$ Ac_{\beta}b_{\gamma}Cb_{\beta}$}&$T_{\uparrow}O:O$  &{\underline{0.14}} &
{$ Ac_{\beta}b_{\gamma}Cb_{\alpha}b_{\beta}Ab_{\gamma}$}&$T_{\uparrow}O:OO:T_{\uparrow}$&{0.60 }&
{$ Ac_{\beta}b_{\gamma}Cb_{\alpha}b_{\beta}Ab_{\gamma}c_{\beta}Bc_{\gamma}$}&$T_{\uparrow}O:OO:T_{\uparrow}O:O$&{2.49} \\
[0.25cm] 
{$\kappa [001]$}&
{$ Ac_{\alpha}c_{\beta}Ba_{\beta}  $} &$OO:O$ &{2.43} &
{$ Ac_{\alpha}c_{\beta}Bc_{\gamma}a_{\beta}Ca_{\alpha}$}&$OO:T_{\uparrow}O:O$&{4.33}&  
{$ Ac_{\alpha}c_{\beta}Bc_{\gamma}a_{\beta}Ca_{\alpha}a_{\gamma}Bc_{\gamma}$}&$OO:T_{\uparrow}O:OO:O$&{2.34}\\
{$\kappa [001]$}& 
{$ Ac_{\alpha}c_{\beta}Bc_{\gamma}$} &$OO:T_{\uparrow}$  &{0.80} &
{$ Ac_{\alpha}c_{\beta}Bc_{\gamma}a_{\beta}Ca_{\gamma}$}&$OO:T_{\uparrow}O:O$&{5.05}&
{$ Ac_{\alpha}c_{\beta}Bc_{\gamma}a_{\beta}Ca_{\alpha}a_{\gamma}Ba_{\beta}$}&$OO:T_{\uparrow}O:OO:T_{\uparrow}$ &{2.56}\\
[0.25cm]
{$\kappa [001]$}& 
{$ Ab_{\alpha}b_{\gamma}Ca_{\gamma} $}&$OO:O$  &{2.20} &
{$ Ab_{\alpha}b_{\gamma}Cb_{\beta}a_{\gamma}Ba_{\alpha} $}&$OO:T_{\uparrow}O:O$&{3.86}&  
{$  Ab_{\alpha}b_{\gamma}Cb_{\beta}a_{\gamma}Ba_{\alpha}a_{\beta}Cb_{\beta}$}&$OO:T_{\uparrow}O:OO:O$&{1.42}\\
{$\kappa [001]$}& 
\underline{$ Ab_{\alpha}b_{\gamma}Cb_{\beta}$}\footnotemark[1]&$OO:T_{\uparrow}$   &{\underline{0.00}} &
{$ Ab_{\alpha}b_{\gamma}Cb_{\beta}a_{\gamma}Ba_{\beta}$}&$OO:T_{\uparrow}O:O$ &{4.30} &
{$Ab_{\alpha}b_{\gamma}Cb_{\beta}a_{\gamma}Ba_{\alpha}a_{\beta}Ca_{\gamma}$}&$OO:T_{\uparrow}O:OO:T_{\uparrow}$&{1.42}\\
[0.25cm] 
\hline
{\T$\kappa [00\bar{1}]$}& 
{$ Ac_{\gamma}a_{\beta}Ba_{\gamma}$  }&$OT_{\downarrow}:O$  &{2.14}&
{$ Ac_{\gamma}a_{\beta}Ba_{\gamma}a_{\alpha}Ca_{\beta}$}&$OT_{\downarrow}:OO:O$ &{3.15}  &
\underline{$  Ac_{\gamma}a_{\beta}Ba_{\gamma}a_{\alpha}Ca_{\beta}c_{\gamma}Bc_{\beta}$}&$OT_{\downarrow}:OO:T_{\downarrow}O:O$&\underline{0.62}\\
{$\kappa [00\bar{1}]$}& 
{$ Ac_{\gamma}a_{\beta}Ba_{\alpha}$   } &$OT_{\downarrow}:O$  &{1.64}&
{$ Ac_{\gamma}a_{\beta}Ba_{\gamma}a_{\alpha}Cc_{\gamma}$ }&$OT_{\downarrow}:OO:T_{\downarrow}$ &{2.25}&
{$ Ac_{\gamma}a_{\beta}Ba_{\gamma}a_{\alpha}Ca_{\beta}c_{\gamma}Bc_{\alpha}$}&$OT_{\downarrow}:OO:OT_{\downarrow}:O$&{0.94}\\
[0.25cm]

{$\kappa [00\bar{1}]$}& 
{$ Ab_{\beta}a_{\gamma}Ca_{\beta}  $} &$OT_{\downarrow}:O$ &{2.15} &
{$ Ab_{\beta}a_{\gamma}Ca_{\beta}a_{\alpha}Ba_{\gamma}$ }   & $OT_{\downarrow}:OO:O$&{2.57} &
{\underline{$ Ab_{\beta}a_{\gamma}Ca_{\beta}a_{\alpha}Ba_{\gamma}b_{\beta}Cb_{\alpha} $}}&$OT_{\downarrow}:OO:T_{\downarrow}O:O$&{\underline{0.44}}\\
{$\kappa [00\bar{1}]$}& 
{$ Ab_{\beta}a_{\gamma}Ca_{\alpha} $} &$OT_{\downarrow}:O$  &{1.24}&
{$ Ab_{\beta}a_{\gamma}Ca_{\beta}a_{\alpha}Bb_{\beta}$ }&$OT_{\downarrow}:OO:T_{\downarrow}$&{1.69} &
{$Ab_{\beta}a_{\gamma}Ca_{\beta}a_{\alpha}Ba_{\gamma}b_{\beta}Cb_{\gamma}$}&$OT_{\downarrow}:OO:T_{\downarrow}O:O$&{1.89}\\
[0.25cm] 
{$\kappa [00\bar{1}]$}& 
{$ Ac_{\gamma}c_{\alpha}Bb_{\gamma} $}  &$OO:T_{\downarrow}$   &{2.36}&
{$ Ac_{\gamma}c_{\alpha}Bc_{\beta}b_{\gamma}Ab_{\alpha}$ }&$OO:OT_{\downarrow}:O$&{1.67}  &
\underline{$ Ac_{\gamma}c_{\alpha}Bc_{\beta}b_{\gamma}Ab_{\beta}b_{\alpha}Cb_{\gamma}$}&$OO:OT_{\downarrow}:OO:O$&\underline{0.00}\\
{$\kappa [00\bar{1}]$}& 
{$ Ac_{\gamma}c_{\alpha}Bc_{\beta}$ } &$OO:O$ &{0.80} &
{$ Ac_{\gamma}c_{\alpha}Bc_{\beta}b_{\gamma}Ab_{\beta}$  }&  $OO:OT_{\downarrow}:O$    &{4.84}  &
{$ Ac_{\gamma}c_{\alpha}Bc_{\beta}b_{\gamma}Ab_{\beta}b_{\alpha}Cc_{\beta}$}&$OO:OT_{\downarrow}:OO:T_{\downarrow}$&{4.62}\\
[0.25cm] 
{$\kappa [00\bar{1}]$}& 
{$ Ab_{\beta}b_{\alpha}Cc_{\beta}$ }&  $OO:T_{\downarrow}$&{2.17} &
\underline{$ Ab_{\beta}b_{\alpha}Cb_{\gamma}c_{\beta}Ac_{\alpha}$ }&$OO:OT_{\downarrow}:O$&{\underline{0.37}}&
{$  Ab_{\beta}b_{\alpha}Cb_{\gamma}c_{\beta}Ac_{\gamma}c_{\alpha}Bc_{\beta}$}&$OO:OT_{\downarrow}:OO:O$&{3.81}\\
{$\kappa [00\bar{1}]$}& 
\underline{$ Ab_{\beta}b_{\alpha}Cb_{\gamma}$\footnotemark[1]} &$OO:O$  &{\underline{0.00}} &
{$ Ab_{\beta}b_{\alpha}Cb_{\gamma}c_{\beta}Ac_{\gamma} $  }&$OO:OT_{\downarrow}:O$&{3.91}&
{$ Ab_{\beta}b_{\alpha}Cb_{\gamma}c_{\beta}Ac_{\gamma}c_{\alpha}Bb_{\gamma}$}&$OO:OT_{\downarrow}:OO:T_{\downarrow}$&{3.91}
\end{tabular}
\end{scriptsize}
\end{ruledtabular}
\footnotetext[1]{The stacking direction 
($[001]$ or $[00\bar{1}]$)
cannot be inferred at the considered thickness.}
\caption{\label{tab:Erel_Al-class}
Stacking sequence and Al coordination
of \textit{unrelaxed} alumina films 
with Al$_{4n-2}$O$_{6n}$ stoichiometry 
and their relative 
total-energy differences 
$E_{\mtext{rel}}$
\textit{after} relaxation.
Notation and grouping 
are the same 
as in Tab.~\ref{tab:Erel_O-class}.
Configurations 
that differ only in their 
surface Al ion are grouped 
together and separated by larger whitespace. 
The coordination 
given for the 
surface Al ion is the one that it would have in the bulk.
The \textit{ab initio} study and comparison permit us to make the 
following set of observations: 
(i)~Although the stable films are
generally of $\kappa$
type, $\alpha$-type films are competitive,
at least for the thinnest 
films; 
(ii)~For the thinner films, both $\kappa [001]$ 
and $\kappa [00\bar{1}]$ orientations yield 
stable and metastable configurations, while for 
the thicker films, only $\kappa [00\bar{1}]$ leads 
to (meta-)stable configurations; 
(iii)~The general trend in 
stability with respect to 
the O stacking 
is the same 
as for 
the Al$_{4n-4}$O$_{6n}$ films
($AC$ more favorable than $AB$)
but with exceptions, in particular,
the stable 
Al$_{10}$O$_{18}$ and Al$_{14}$O$_{24}$ 
configurations originate from structures with 
$AB$ stacking in the first two O layers. 
}
\end{table*}

\begin{table}
\begin{ruledtabular}
\begin{scriptsize}
\begin{tabular}{l|llr}
alumina& 
alumina& coord. of & $E_{\mtext{rel}}$\\
\B group  &
stacking& Al ions& (eV/cell)\B\\
\hline
&\mc{3}{c}{\T\underline{
Al$_8$O$_{12}$ films}}\T\\[0.1cm]
{$\alpha$}& 
{$Ac_3c_2Bc_1c_3$}& $OO:OO$  &{2.46}\\
{$\alpha$}& 
{$Ab_2b_3Cb_1b_2 $}&$OO:OO$  & {1.92}\B\\
\hline
{\T$\kappa [001]$}&
{$Ab_{\gamma}c_{\beta}Bc_{\alpha}c_{\gamma}$}&
$T_{\uparrow}O:OO$ &{3.17} \\
{$\kappa [001]$}& 
{$Ac_{\beta}b_{\gamma}Cb_{\alpha}b_{\beta}$ }&
$T_{\uparrow}O:OO$ &{2.57} \\
[0.25cm] 
{$\kappa [001]$}& 
{$Ac_{\alpha}c_{\beta}Bc_{\gamma}a_{\beta}$ } &
$OO:T_{\uparrow}O$  &{4.07} \\
{$\kappa [001]$}& 
{$Ab_{\alpha}b_{\gamma}Cb_{\beta}a_{\gamma}$} & 
$OO:T_{\uparrow}O$ &{2.06\B} \\
\hline
\T{$\kappa [00\bar{1}]$}& 
{$ Ac_{\gamma}a_{\beta}Ba_{\gamma}a_{\alpha}$  }&
$OT_{\downarrow}:OO$  &  {2.08} \\
{$\kappa [00\bar{1}]$}& 
\underline{$ Ab_{\beta}a_{\gamma}Ca_{\alpha}a_{\beta}$}& 
$OT_{\downarrow}:OO$ &  {\underline{0.00}}\\
[0.25cm] 
{$\kappa [00\bar{1}]$}& 
{$ Ac_{\gamma}c_{\alpha}Bc_{\beta}b_{\gamma}$  }& 
$OO:OT_{\downarrow}$  &  {2.90} \\
{$\kappa [00\bar{1}]$}& 
{$ Ab_{\beta}b_{\alpha}Cb_{\gamma}c_{\beta}$ } & 
$OO:OT_{\downarrow}$    & {1.80}\B \\
\hline
&\mc{3}{c}{\T \underline{
Al$_{12}$O$_{18}$ 
films}}\\[0.15cm]
{$\alpha$}& 
{$Ac_3c_2Bc_1c_3Ac_2c_1$}           & $OO:OO:OO$
&{2.88}  \\
{$\alpha$}& 
{$Ab_2b_3Cb_1b_2Ab_3b_1$}            & $OO:OO:OO$
&{3.93}\B  \\
\hline    
{\T$\kappa [001]$}& 
{$Ab_{\gamma}c_{\beta}Bc_{\alpha}c_{\gamma}Ac_{\beta}b_{\gamma}$ }& $T_{\uparrow}O:OO:T_{\uparrow}O$
&{1.42}  \\
{$\kappa [001]$}& 
{$Ac_{\beta}b_{\gamma}Cb_{\alpha}b_{\beta}Ab_{\gamma}c_{\beta} $}& $T_{\uparrow}O:OO:T_{\uparrow}O$
&{1.16}  \\
[0.25cm] 
{$\kappa [001]$}& 
{$Ac_{\alpha}c_{\beta}Bc_{\gamma}a_{\beta}Ca_{\alpha}a_{\gamma}$ }& $OO:T_{\uparrow}O:OO$
&{2.78}  \\
{$\kappa [001]$}& 
{$Ab_{\alpha}b_{\gamma}Cb_{\beta}a_{\gamma}Ba_{\alpha}a_{\beta} $  }&$OO:T_{\uparrow}O:OO$
&{1.48}\B\\
\hline
\T{$\kappa [00\bar{1}]$}& 
{$ Ac_{\gamma}a_{\beta}Ba_{\gamma}a_{\alpha}Ca_{\beta}c_{\gamma}$  }&$OT_{\downarrow}:OO:OT_{\downarrow}$
&{0.49}  \\
{$\kappa [00\bar{1}]$}& 
\underline{$ Ab_{\beta}a_{\gamma}Ca_{\beta}a_{\alpha}Ba_{\gamma}b_{\beta}$}&$OT_{\downarrow}:OO:OT_{\downarrow}$
&{\underline{0.00}}  \\
[0.25cm] 
\T{$\kappa [00\bar{1}]$}&
{$ Ac_{\gamma}c_{\alpha}Bc_{\beta}b_{\gamma}Ab_{\beta}b_{\alpha}$ }&$OO:OT_{\downarrow}:OO$
&{1.72}\\
{$\kappa [00\bar{1}]$}& 
{$ Ab_{\beta}b_{\alpha}Cb_{\gamma}c_{\beta}Ac_{\gamma}c_{\alpha}$ }& $OO:OT_{\downarrow}:OO$
&{0.52} 
\end{tabular}
\end{scriptsize}
\end{ruledtabular}
\caption{\label{tab:Erel_AlAl-class}
Stacking sequence and Al coordination
of the \textit{unrelaxed} 
alumina films 
with Al$_{4n}$O$_{6n}$ stoichiometry 
and their relative 
total-energy differences 
$E_{\mtext{rel}}$
\textit{after} relaxation.  Notation and grouping 
are the same as in Tab.~\ref{tab:Erel_O-class}. 
The \textit{ab initio} study and comparison permit 
a number of observations that are similar to those we made for 
the Al$_{4n-4}$O$_{6n}$ films (Tab.~\ref{tab:Erel_O-class}). 
However, the 
Al$_{4n}$O$_{6n}$ 
films are thermodynamically unstable,
see Sec.~\ref{sec:Thermodynamics}.
}
\end{table}

\section{Results I: Energetical  and Thermodynamical Stability\label{sec:Energetics}}
In Tables~\ref{tab:Erel_O-class}--\ref{tab:Erel_AlAl-class} 
we list  all two, three, and four O 
layersthick (initial)
alumina-film configurations 
that are consistent with the bulk $\alpha$- or \Kalumina\
structure. 
The configurations are grouped according to their 
stoichiometry, film thickness, and the phase and 
orientation of the alumina bulk structures from 
which they are derived. 

For each configuration, we list the stacking 
sequence and the coordination of the Al 
ions
of the alumina film before relaxation, together 
with the calculated total energies after relaxation. 
These energies ($E_{\mtext{rel}}$) are given relative 
to the energy $E_0$ of the energetically lowest lying 
configuration of the same thickness and stoichiometry 
class, $E_{\mtext{rel}}=E-E_0$.

\subsection{Energetics and metastability\label{sec:Metastability}} 
For each thickness and stoichiometry class
we use the total energy of the energetically most-favorable
configuration (the candidate for the stable thin film structure) 
to define a zero-point of relative energy differences,
$E_{\mtext{rel}}=0$.
We stress, however, that 
bulk alumina also exists in a number 
of metastable phases.
We may therefore also expect metastable configurations
among the thin films.

In our calculation for bulk alumina, we find that 
the DFT energy difference between the $\kappa$ 
and $\alpha$ phases is 
$\Delta_{\alpha\kappa}\sim 0.7$~eV/\alumina. 
We use this quantity as an indicative measure of the 
metastability of the alumina films 
and define
$E_{\mtext{meta}}= 2 L_{\mtext{O}} \Delta_{\alpha\kappa}$, 
where $L_{\mtext{O}}$ is the number of O layers in the alumina 
film.  
For stoichiometric films, $2 L_{\mtext{O}}$ is equal to 
the number of stoichiometric Al$_2$O$_3$ units in the film. 
For non-stoichiometric films, it will give an approximate 
measure of the number of stoichiometric units. 
We then consider configurations with 
$E_{\mtext{rel}}\gtrsim E_{\mtext{meta}}$
as unstable and configurations
with $0\leq E_{\mtext{rel}}\lesssim E_{\mtext{meta}}$
as potentially metastable.
Whether or not they are truly
metastable cannot, however, 
be inferred from 
our calculations.
Such an analysis would require a calculation of the nature of
vibrational excitations and 
is much beyond the present 
investigation.

For the two, three, and four O layers thick films 
the criterion is 
$E_{\mtext{meta}}=0.28$~eV,
$0.42$~eV, and 
$0.56$~eV, respectively.
In Tables~\ref{tab:Erel_O-class}--\ref{tab:Erel_AlAl-class}, 
the configurations that are
stable or potentially metastable are underlined.
For the Al$_{4n-4}$O$_{6n}$ stoichiometry class, 
there are at most two potentially 
metastable configurations for each alumina film thickness. 
For Al$_{4n}$O$_{6n}$, no potentially metastable configurations
are found. 
Among the Al$_{4n-2}$O$_{6n}$ configurations,
metastable films are more common.
However, their number decreases as the 
film thickness increases.

\begin{figure*}
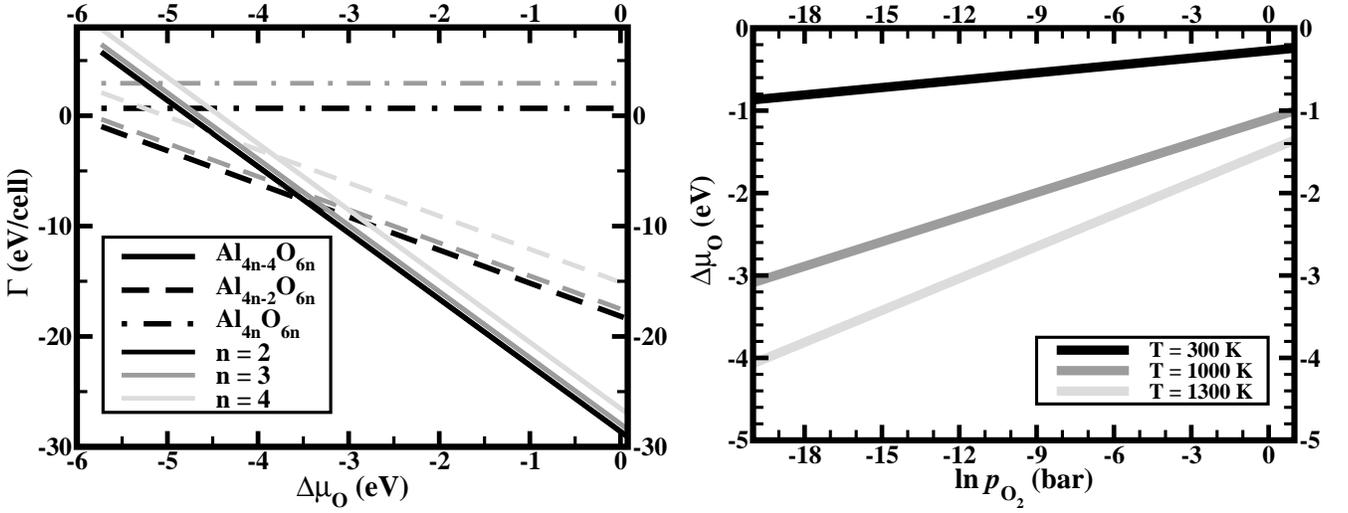

\begin{tabular}{lcr}
\epsfig{file=fig3a.eps,width=8.8cm}&&
\epsfig{file=fig3b.eps,width=8.4cm}
\end{tabular}
\caption{\label{fig:GibbsEquilibrium}
Thermodynamic stability of thin-film alumina 
with stoichiometrically different compositions
and different thicknesses on the TiC(111) substrate
in equilibrium with an O$_2$ environment.
The left panel shows the Gibbs free energy differences $\Gamma$
(Eq.~\ref{Eq:Gibbs_0}) per unit cell of TiC/thin-film alumina 
as a function of the O chemical potential
$\Delta\mu_{\mtext{O}}=\mu_{\mtext{O}}-\frac{1}{2}\epsilon_{\mtext{O$_2$}}$
(Eq.~\ref{Eq:mu_O})
for all three considered thicknesses and stoichiometric compositions.
The left end of each line is defined by the physically
allowed range (fcc-Al condensation) of the O chemical potential
(Eq.~\ref{Eq:mu-range}).
For all thicknesses the alumina films with Al$_{4n-4}$O$_{6n}$ stoichiometry
(solid lines) are stable at medium to high O chemical potential, 
whereas films with Al$_{4n-2}$O$_{6n}$ stoichiometry (dashed lines)
are stable at low O chemical potential ($\Delta\mu_{\mtext{O}}<-3.5$~eV).
The alumina films with Al$_{4n}$O$_{6n}$ stoichiometry (dashed-dotted lines) 
are not stable at any allowed value of the O chemical potential.
The right panel shows the O chemical potential $\Delta\mu_{\mtext{O}}$
as a function of partial O$_2$ pressure for three different temperatures.
We find that a value of $\Delta\mu_{\mtext{O}}<-3.5$~eV
at a temperature $T=1300$~K corresponds to an
O$_2$ pressure of  $P_{\mtext{O$_2$}}\sim10^{-15}$ bar.
}
\end{figure*}

\subsection{Thermodynamical stability\label{sec:Thermodynamics}}
Figure~\ref{fig:GibbsEquilibrium} shows 
our calculated values of $\Gamma$
for the energetically most favorable  
configurations of each 
considered thickness and stoichiometry class.
Corresponding values for the 
potentially metastable configurations
can be obtained by adding 
the relative energies of
Tables~\ref{tab:Erel_O-class}--\ref{tab:Erel_AlAl-class}.

In the physically allowed range of the O chemical potential
$\mu_{\mtext{O}}$ (Eq.~\ref{Eq:mu-range}),
we find that the stable film belongs to either the 
Al$_{4n-2}$O$_{6n}$ or the Al$_{4n-4}$O$_{6n}$ 
stoichiometry class, independent of the 
thickness.
The Al$_{4n-4}$O$_{6n}$ films are stabilized for 
$\Delta\mu_{\mtext{O}}\geq -3.5$~eV,
whereas for  $\Delta\mu_{\mtext{O}}\leq -3.5$~eV, the 
Al$_{4n-2}$O$_{6n}$ films are stable.
These findings apply when the films are in
equilibrium with an O$_2$ atmosphere.

In the right panel of  Fig.~\ref{fig:GibbsEquilibrium}
we show the relation between the O chemical potential
and the O$_2$ pressure at several temperatures.
We find that an O chemical potential of 
$\Delta\mu_{\mtext{O}}\geq -3.5$~eV
can only be reached for relatively high temperatures 
and extremely low O$_2$ pressures 
($T\sim 1300$~K, $p_{\mtext{O}_2}\sim10^{-15}$~bar (UHV)).

At normal conditions, 
equilibrium thermodynamics
therefore predicts the 
observation of 
Al$_{4n-4}$O$_{6n}$ films.\cite{ref:GammaWithDelta}
We note that this result is not
in contradiction with 
studies on the 
$\alpha$-Al$_2$O$_3 (0001)$ 
or 
$\kappa$-Al$_2$O$_3 \{001\}$ 
surfaces,
which predict Al 
termination.\cite{ref:Carlo_Surfaces,ref:alphaSurfaces_Thermodynamics} 
In our study, the metallic substrate
can take or give away charge via the interface,
so that the polarity argument
that rules out an 
O-terminated surface of
a pure alumina slab
cannot be applied to TiC/thin-film alumina.

\subsection{Trends in phase content, orientation, and preferred stacking
\label{sec:InterfacialOrientation}}
It is clear that a detailed analysis of 
the trends in phase content, orientation, 
and preferred stacking
of the alumina thin films 
must be based on the relaxed configurations.
We find that a classification
of thin-film candidate configurations 
that is based on the unrelaxed 
structures is dangerous.

The stable and metastable alumina films
are in general obtained from truncated
TiC/\Kalumina$[00\bar{1}]$ 
interface configurations.
This would be in agreement with the experimental 
results that growth of \Kalumina\ is 
preferred over \Aalumina\ on clean 
TiC(111) substrate.\cite{ref:TiX-Al2O3_Coatings}
However, truncated TiC/\Aalumina\ 
and TiC/\Kalumina$[001]$ 
configurations are (meta-)stable in the 
case of the thinner films with 
Al$_{4n-2}$O$_{6n}$ stoichiometry.

Comparing configurations that are derived 
from structures that differ only by 
a reflection about the $yz$-plane, we notice 
that, generally, the unrelaxed structure with 
$AC$ stacking in the bottom two O layers yield more 
favorable configurations than the structures 
with $AB$ stacking.  There are, however, exceptions;
in particular the energetically most favorable Al$_{10}$O$_{18}$
film and each one of the potentially metastable
Al$_{10}$O$_{18}$ and Al$_{14}$O$_{24}$ films
posses an $AB$ stacking in the bottom two O layers.
We also note that, in general, the detailed stacking 
sequence of the energetically most favorable configurations 
varies strongly with the film thickness. 

In summary, although there are some general stability trends
that can be inferred from the phase content, orientation, 
and stacking of the unrelaxed thin-film configurations, 
there are also several noticeable exceptions. 
In particular the 
Al$_{4n-2}$O$_{6n}$ 
films tend to break the rules.


\begin{figure}
\begin{tabular}{c}
\epsfig{file=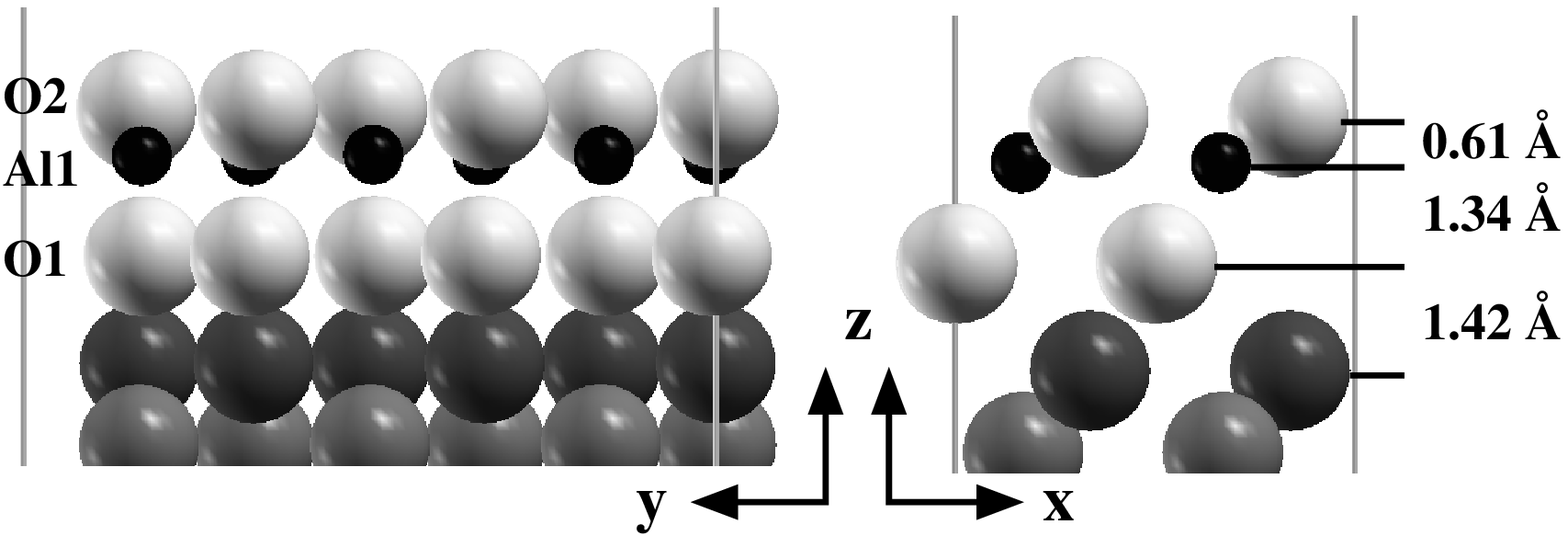,width=8.4cm}\\
\epsfig{file=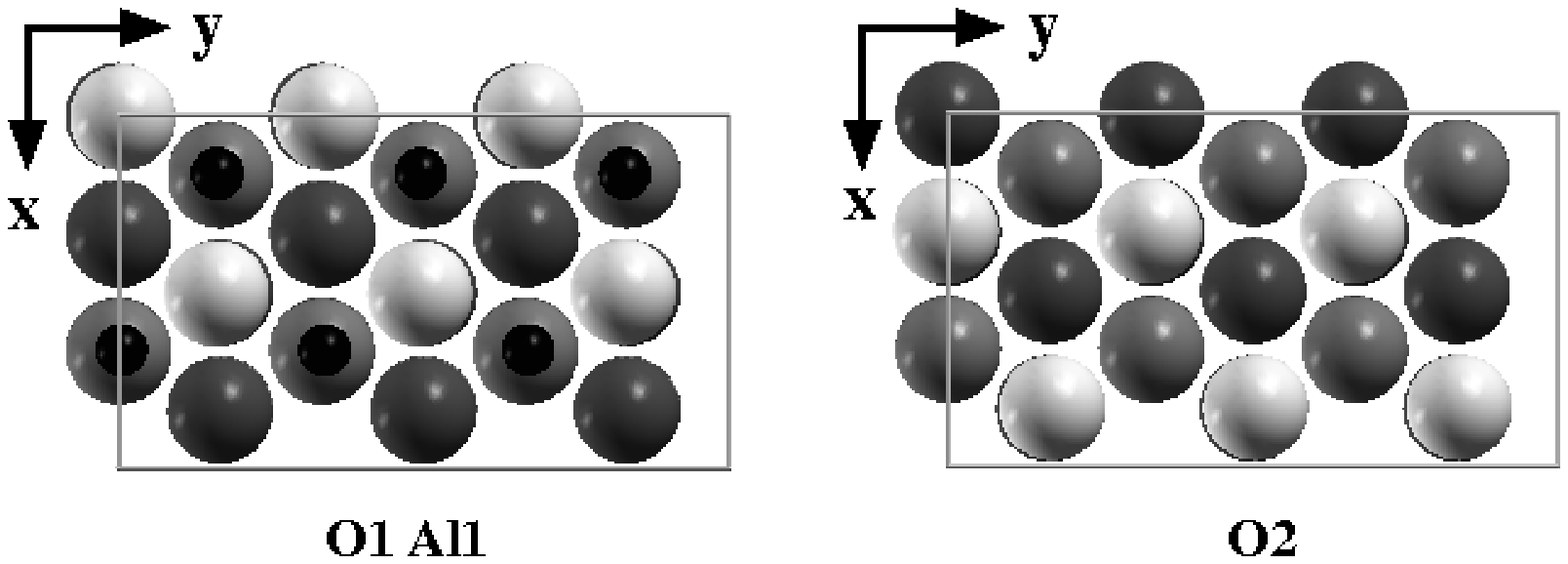,width=8.4cm}\\
\end{tabular}
\caption{
\label{fig:II-Al}
Atomic structure of the stable 
Al$_6$O$_{12}$ 
film.
The color coding is: 
Dark gray = Ti, light gray = C, light = O, and black = Al.
The top panel shows the projected side views along $[100]$ and $[010]$
including interlayer distances.
The bottom panel shows the top views on 
the atomic layers [as defined in the top panel].
The Al coordination
is $OOO$
($O$: octahedral, $T$: tetrahedral).
Note that the film is 
O terminated after 
relaxation.}
\end{figure}

\begin{figure}
\begin{tabular}{c}
\epsfig{file=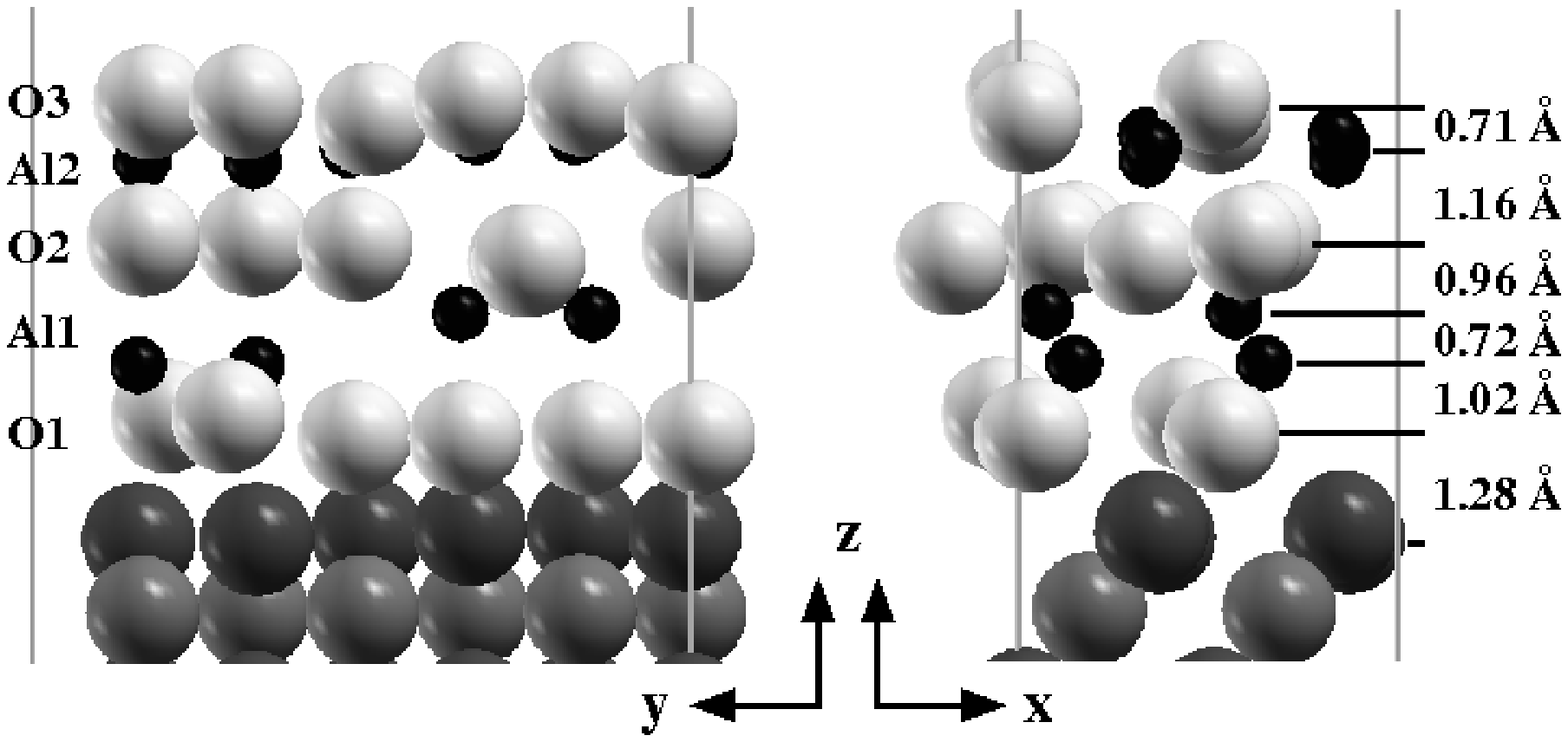,width=8.4cm}\\
\textbf{(a)}\\
\epsfig{file=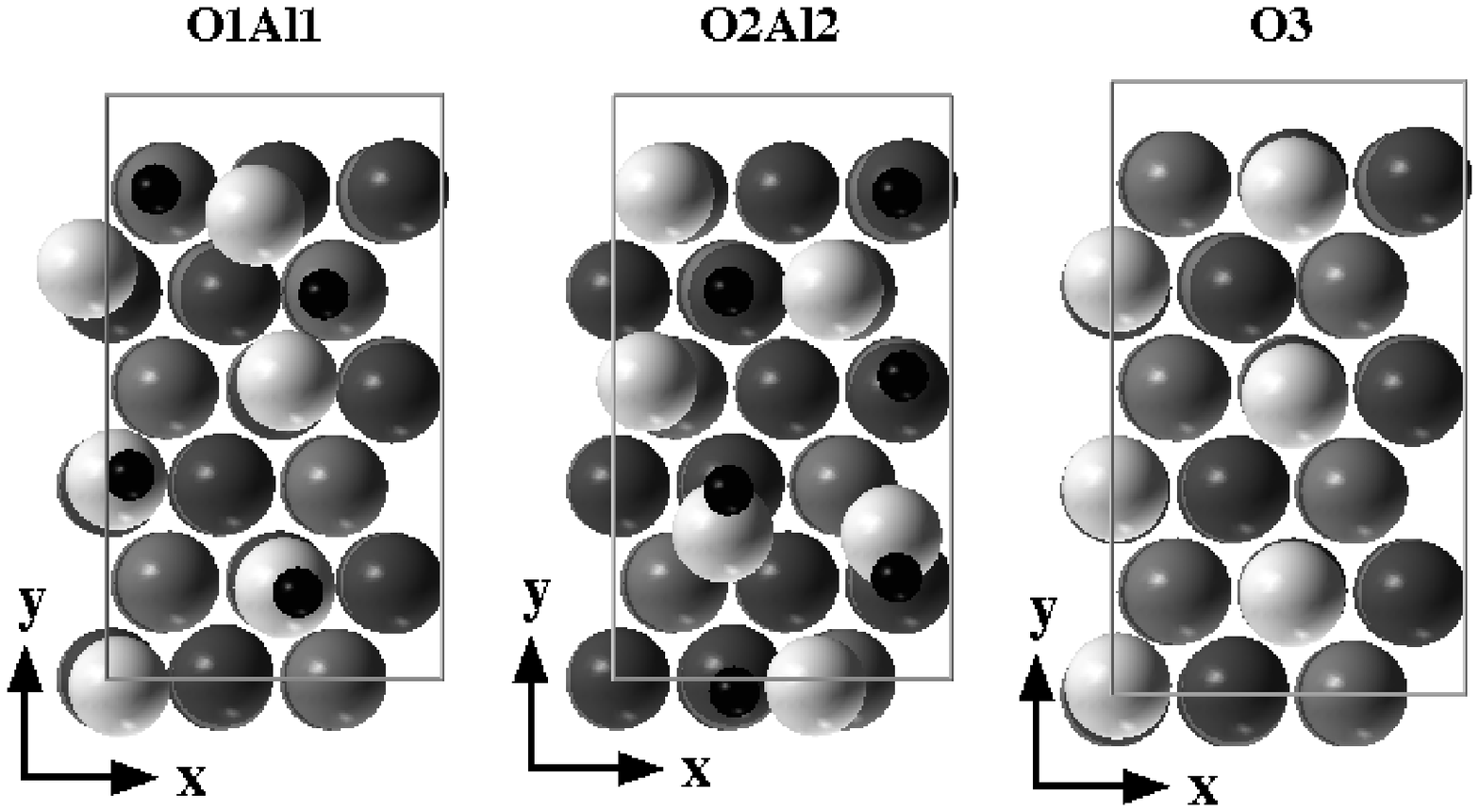,width=8.4cm}\\
\textbf{(b)}
\end{tabular}
\caption{
\label{fig:III-Al}
Atomic structure of the stable 
Al$_{10}$O$_{18}$ 
film (color coding as in Fig.~\ref{fig:II-Al}).
The film is  O terminated after relaxation.
The Al coordination is 
$T_{\downarrow}T_{\uparrow}:T_{\downarrow}OO$
($O$: octahedral, $T$: tetrahedral, the 
arrows indicate the direction in which the 
tetrahedra point, different Al layers are separated by '$:$').
Note that tetrahedral coordination dominates
and that both Al pairs 
in the bottom layer are 
tetrahedrally coordinated, with 
tetrahedra pointing in opposite
directions ($T_{\uparrow}T_{\downarrow}$).
}
\end{figure}

\begin{figure*}
\begin{tabular}{c|c}
\epsfig{file=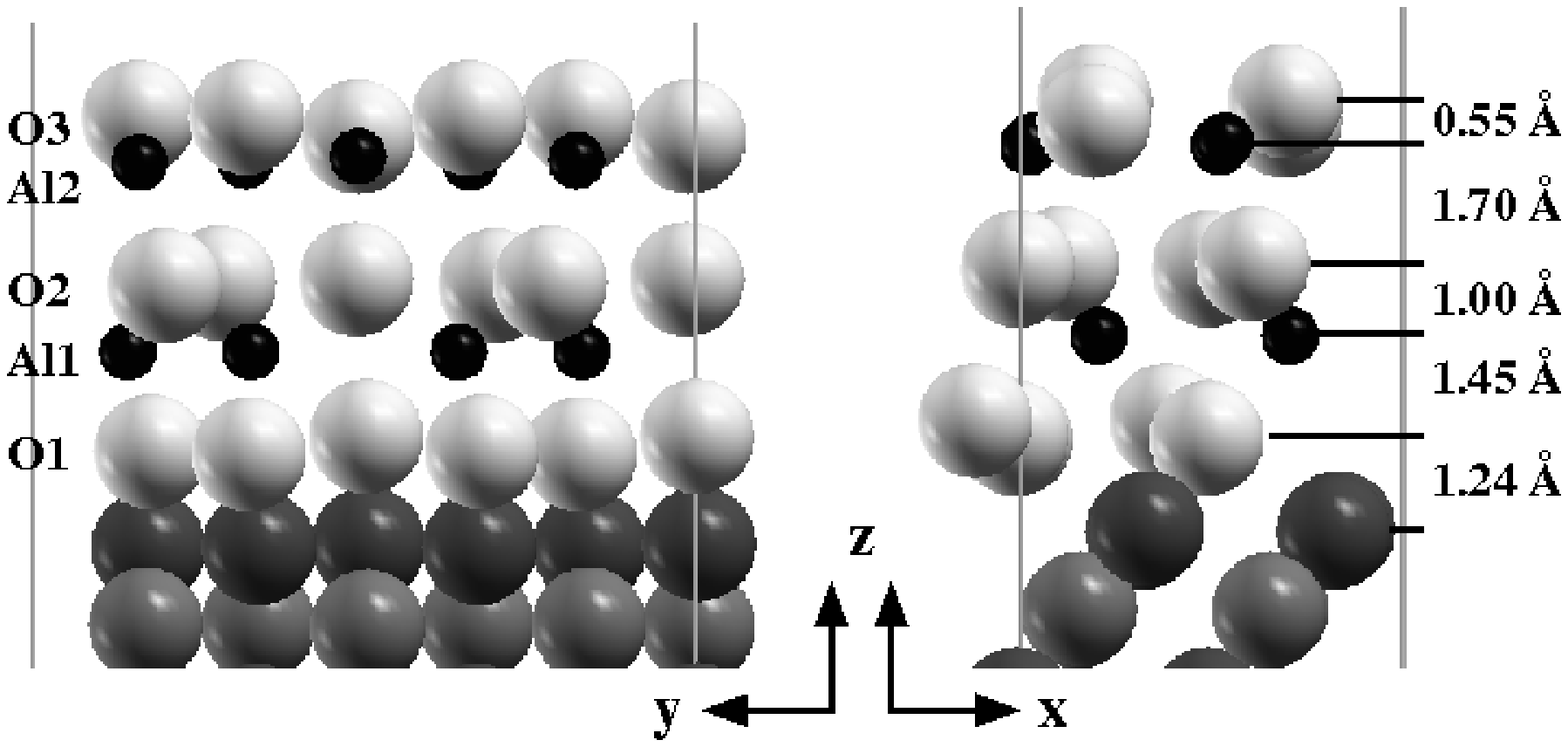,width=8.4cm}&
\epsfig{file=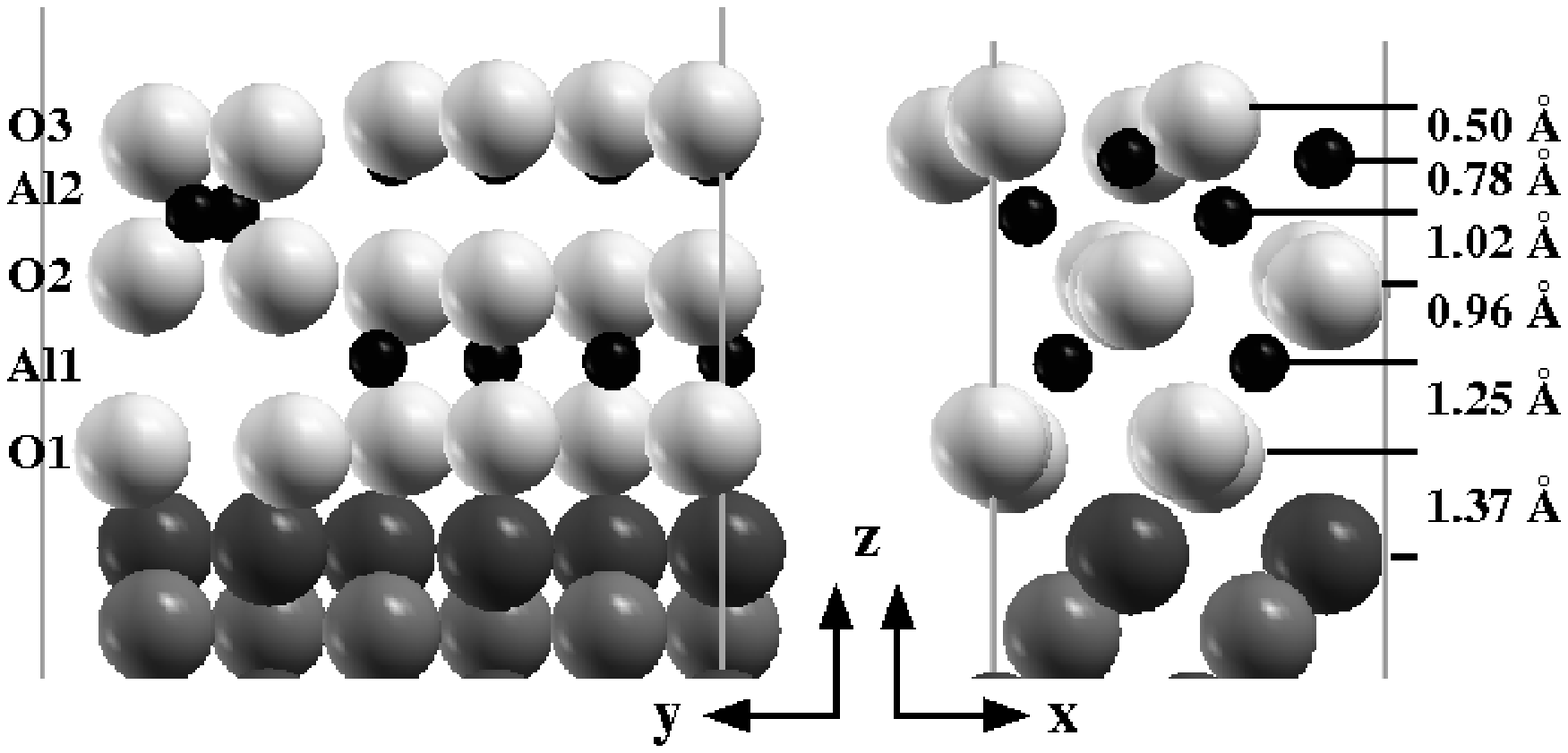,width=8.4cm}\\
\epsfig{file=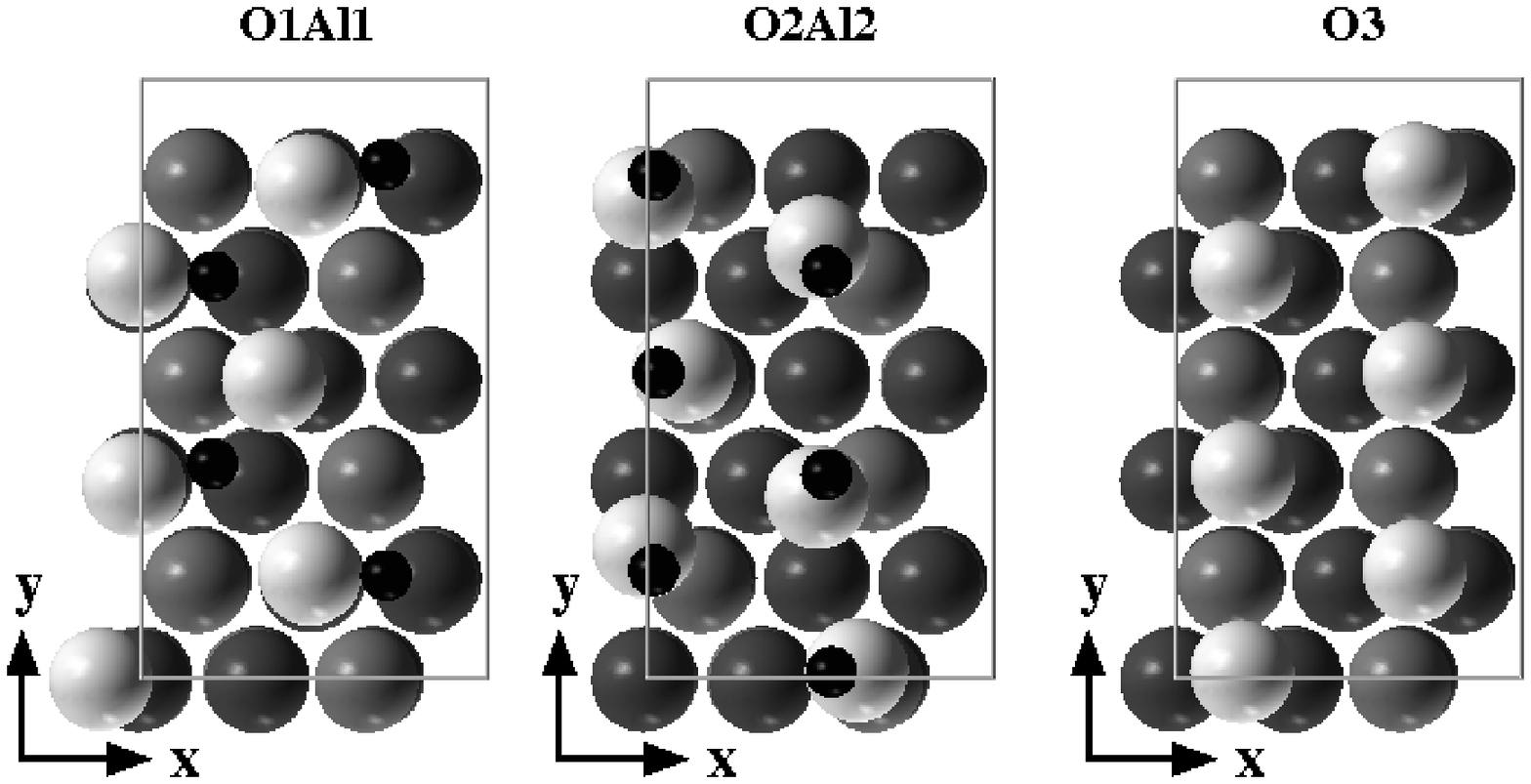,width=8.4cm}&
\epsfig{file=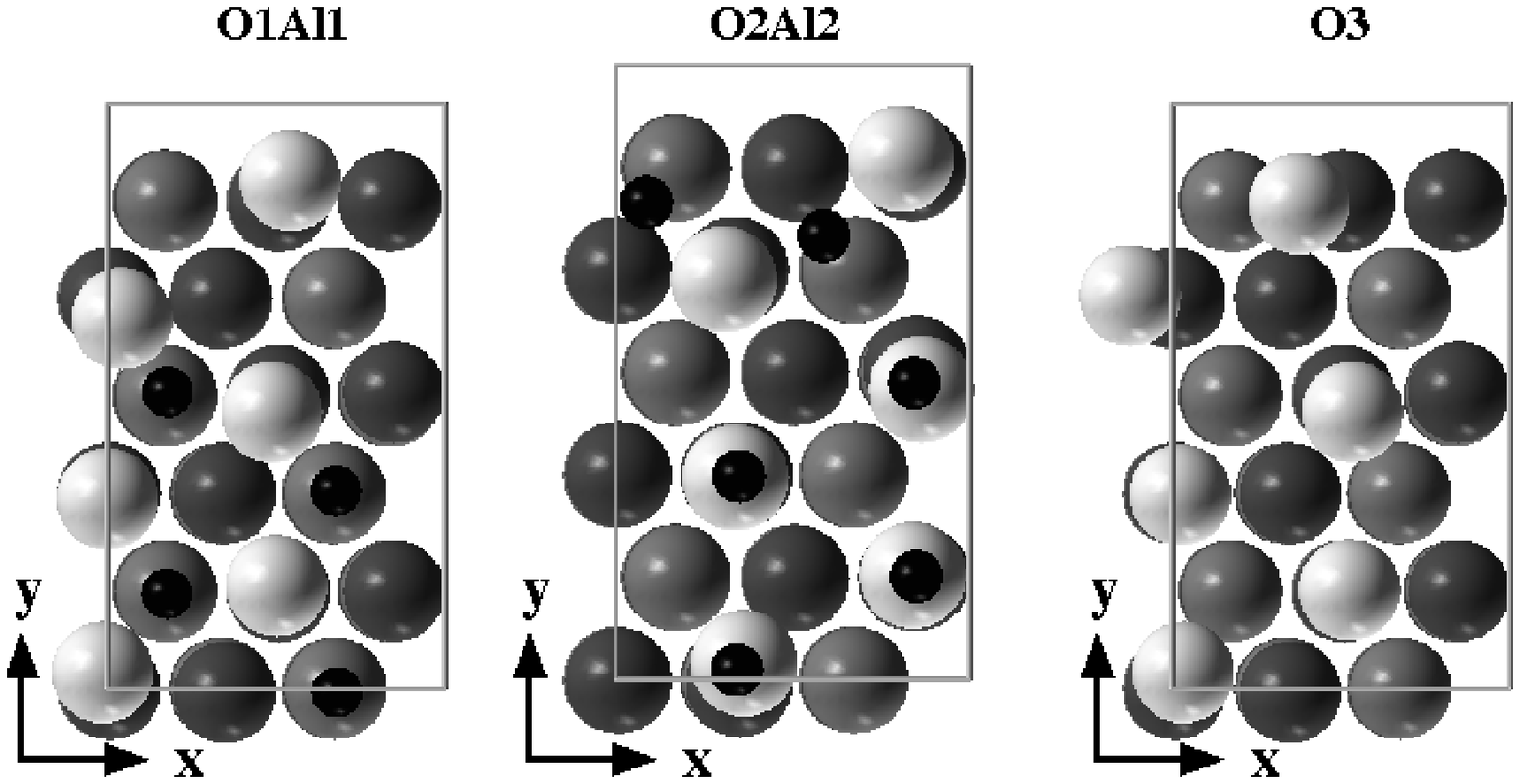,width=8.4cm}\\
\textbf{(a)}&\textbf{(b)}
\end{tabular}
\caption{
\label{fig:Al_metaIII}
Atomic structure of the two potentially metastable 
Al$_{10}$O$_{18}$  films. 
Color coding and notation are as in Fig.~\ref{fig:II-Al}.
The Al coordinations  are 
(a)~$OO:T_{\downarrow}T_{\downarrow}T_{\downarrow}$
and
(b)~$OO:T_{\downarrow}T_{\downarrow}O$
($O$: octahedral, $T$: tetrahedral, 
the arrows indicate the direction in which the 
tetrahedra point, different Al layers are separated by '$:$').
}
\end{figure*}

\begin{figure}
\begin{tabular}{c}
\epsfig{file=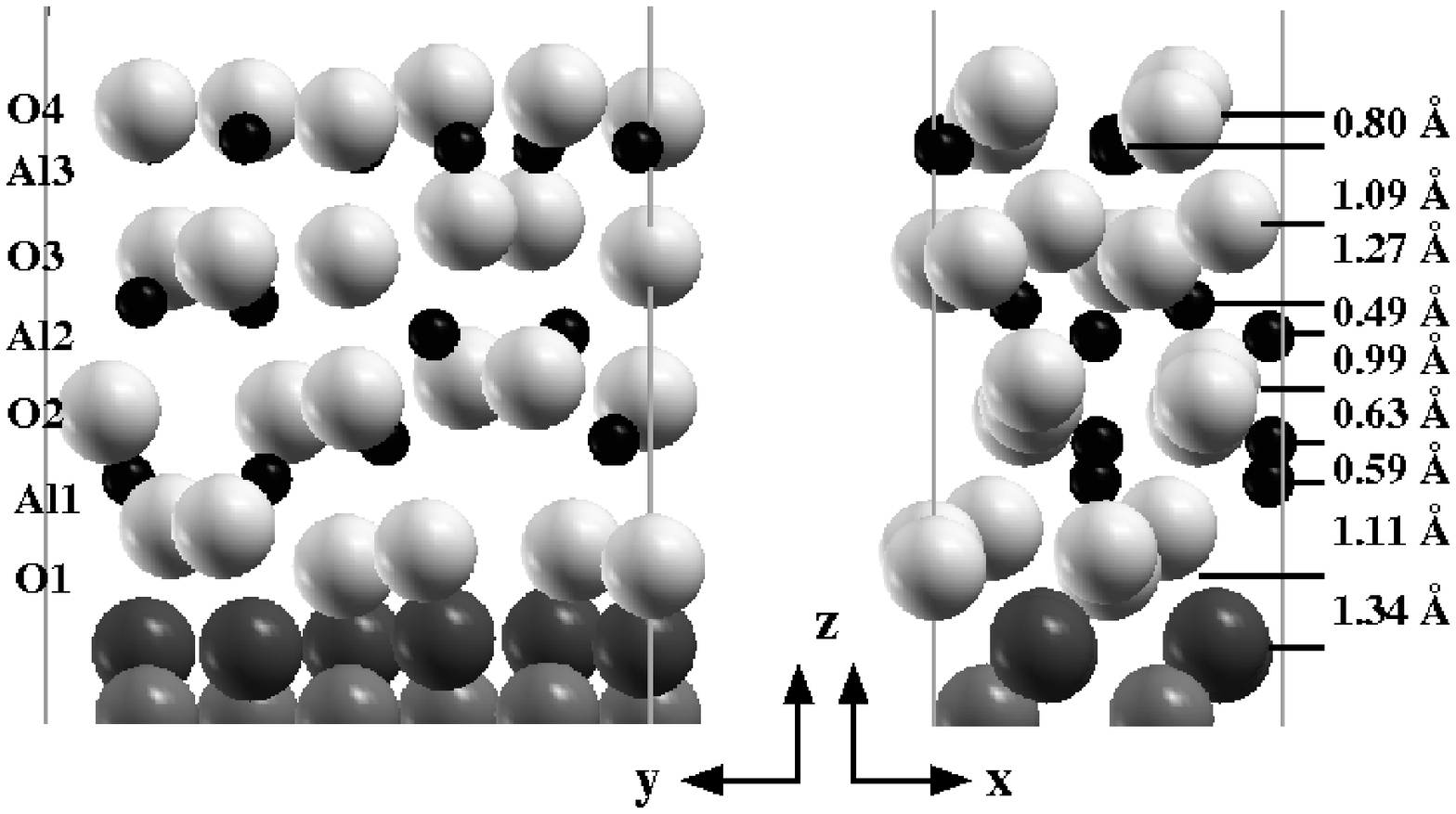,width=8.4cm}\\
\epsfig{file=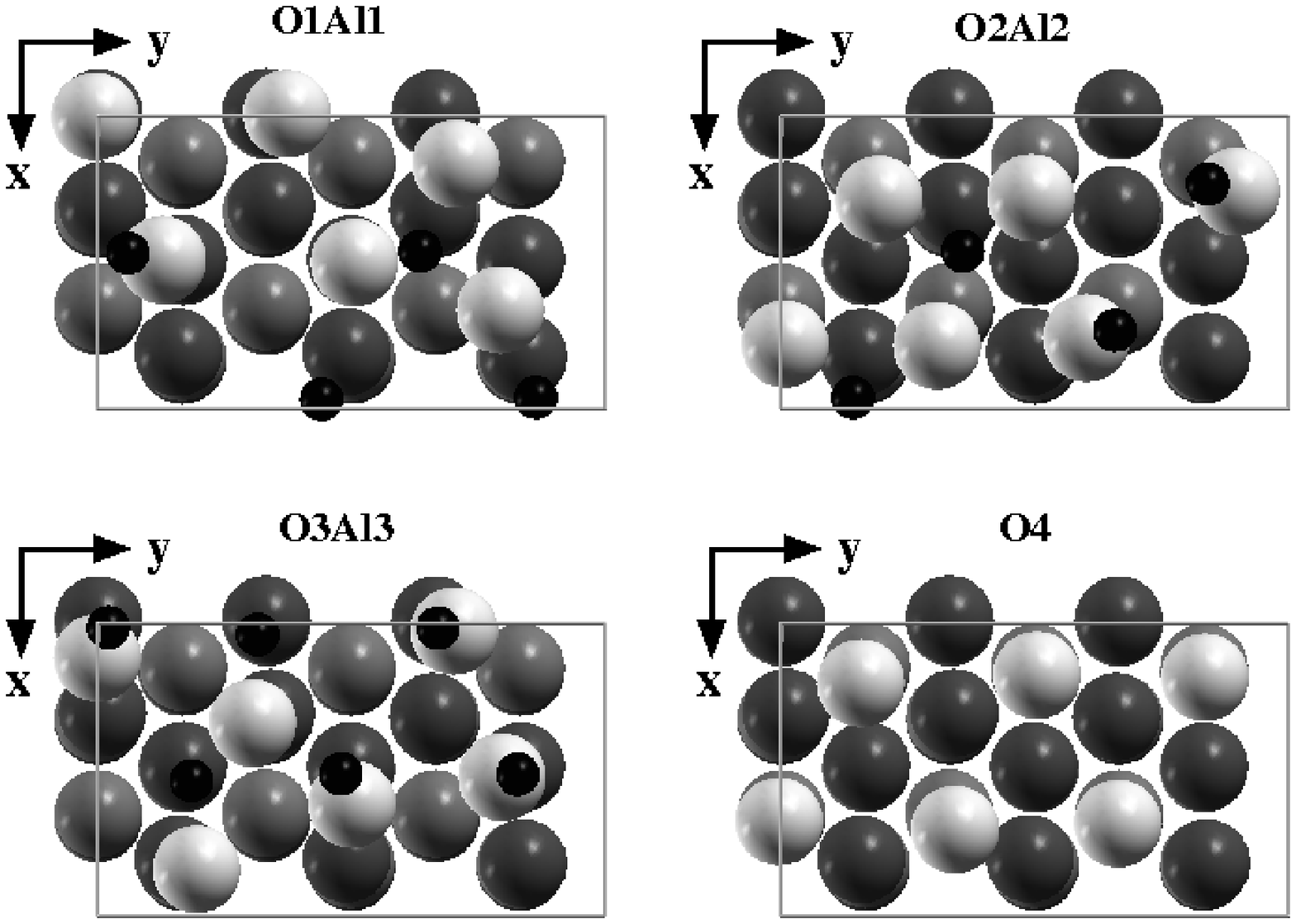,width=8.4cm}\\
\end{tabular}
\caption{
\label{fig:IV-Al}
Atomic structure of the stable 
Al$_{14}$O$_{24}$ 
film (color coding and notation as in Fig.~\ref{fig:II-Al}).
The film is O terminated after relaxation.
The Al coordination is 
$OT_{\downarrow}:T_{\uparrow}T_{\downarrow}:OT_{\downarrow}T_{\downarrow}$
($O$: octahedral, $T$: tetrahedral, 
the  arrows indicate the direction in which the tetrahedra point,
different Al layers are separated by '$:$').
As in the most favorable Al$_{10}$O$_{18}$ film, 
there  is a layer in which both Al pairs are 
tetrahedrally coordinated, with tetrahedra 
pointing in opposite directions ($T_{\downarrow}T_{\uparrow}$).
Here, they are located in the second Al layer. 
}
\end{figure}

\begin{figure*}
\begin{tabular}{c|c}
\epsfig{file=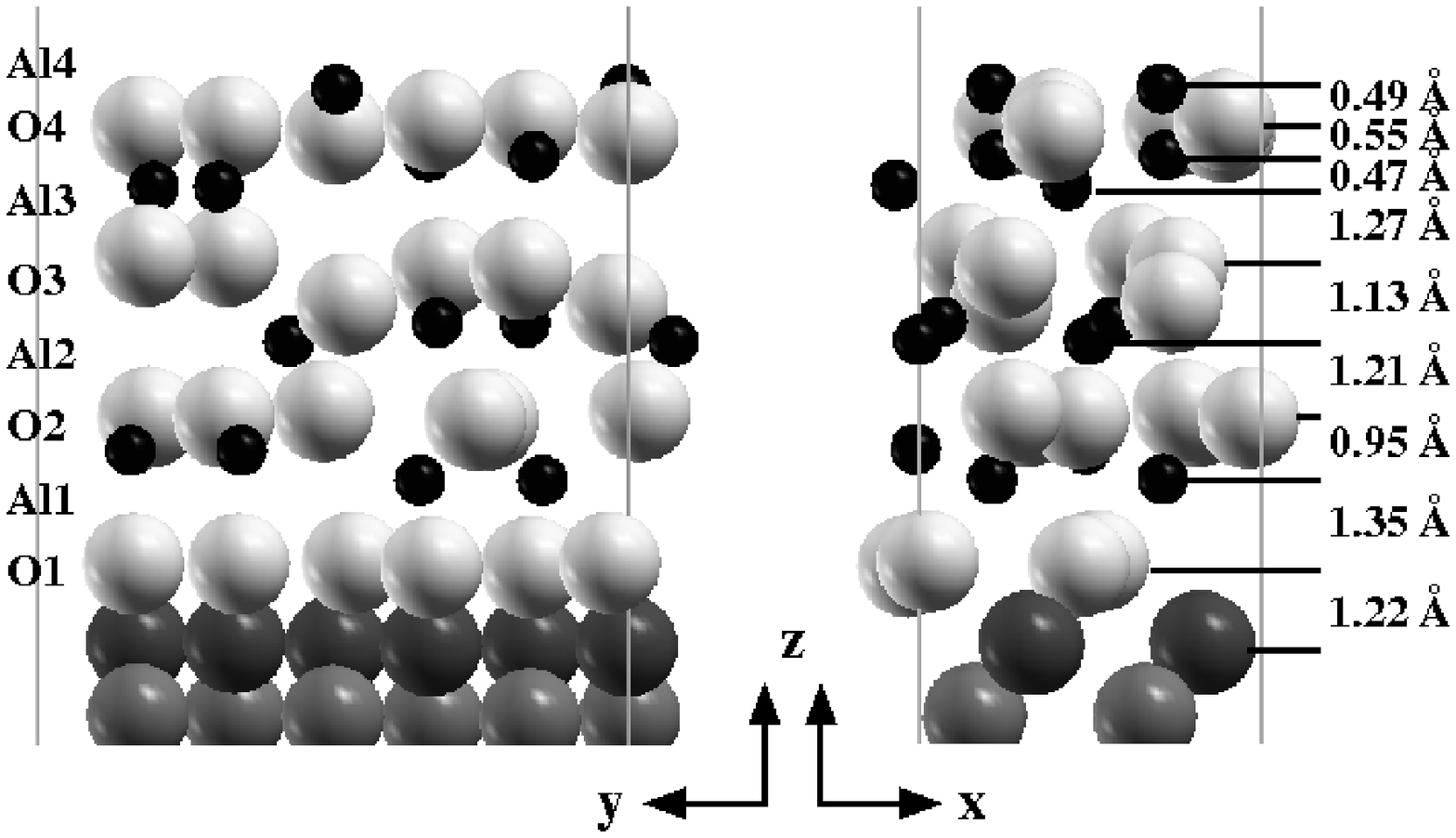,width=8.4cm}&
\epsfig{file=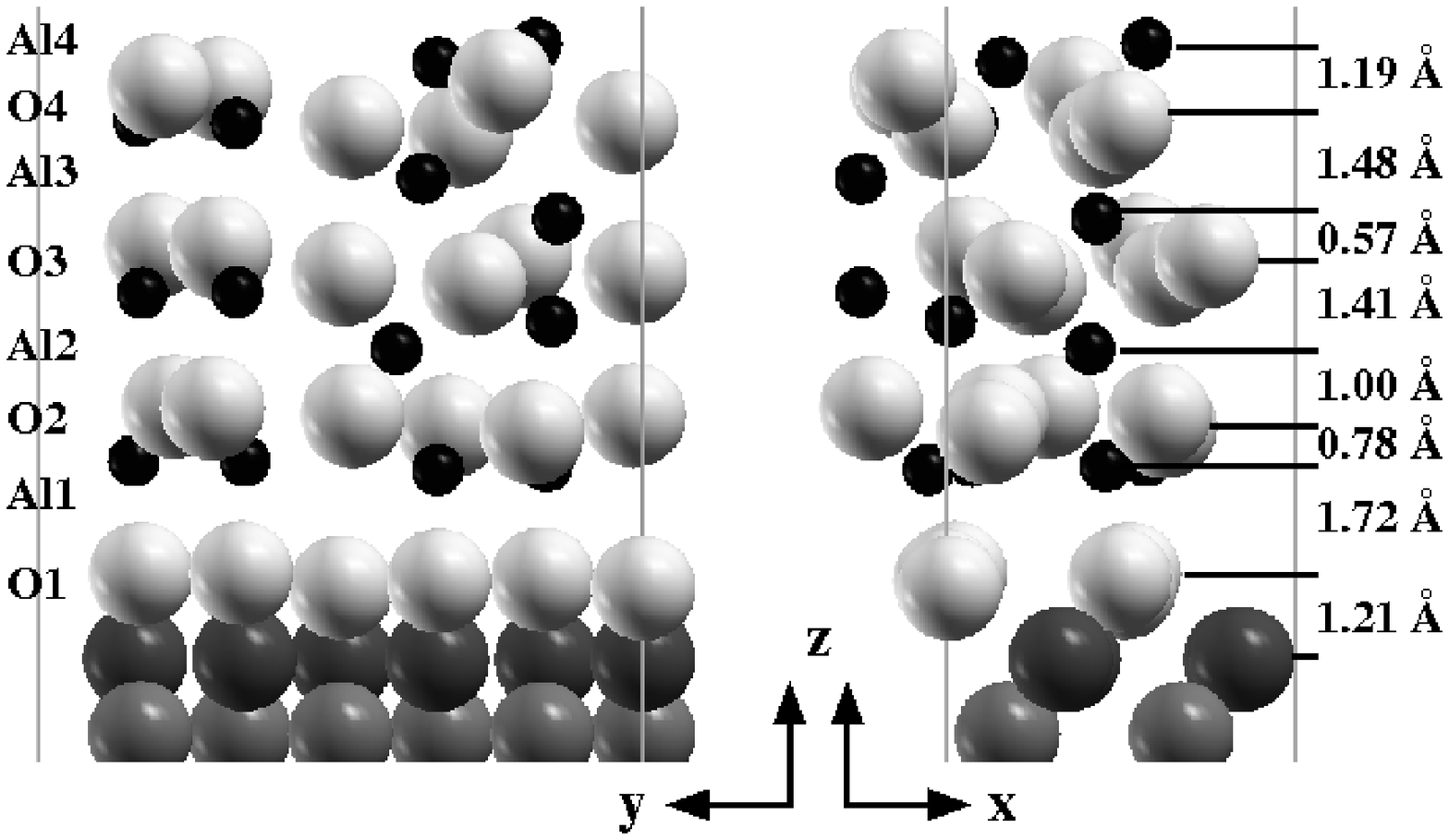,width=8.4cm}\\
\epsfig{file=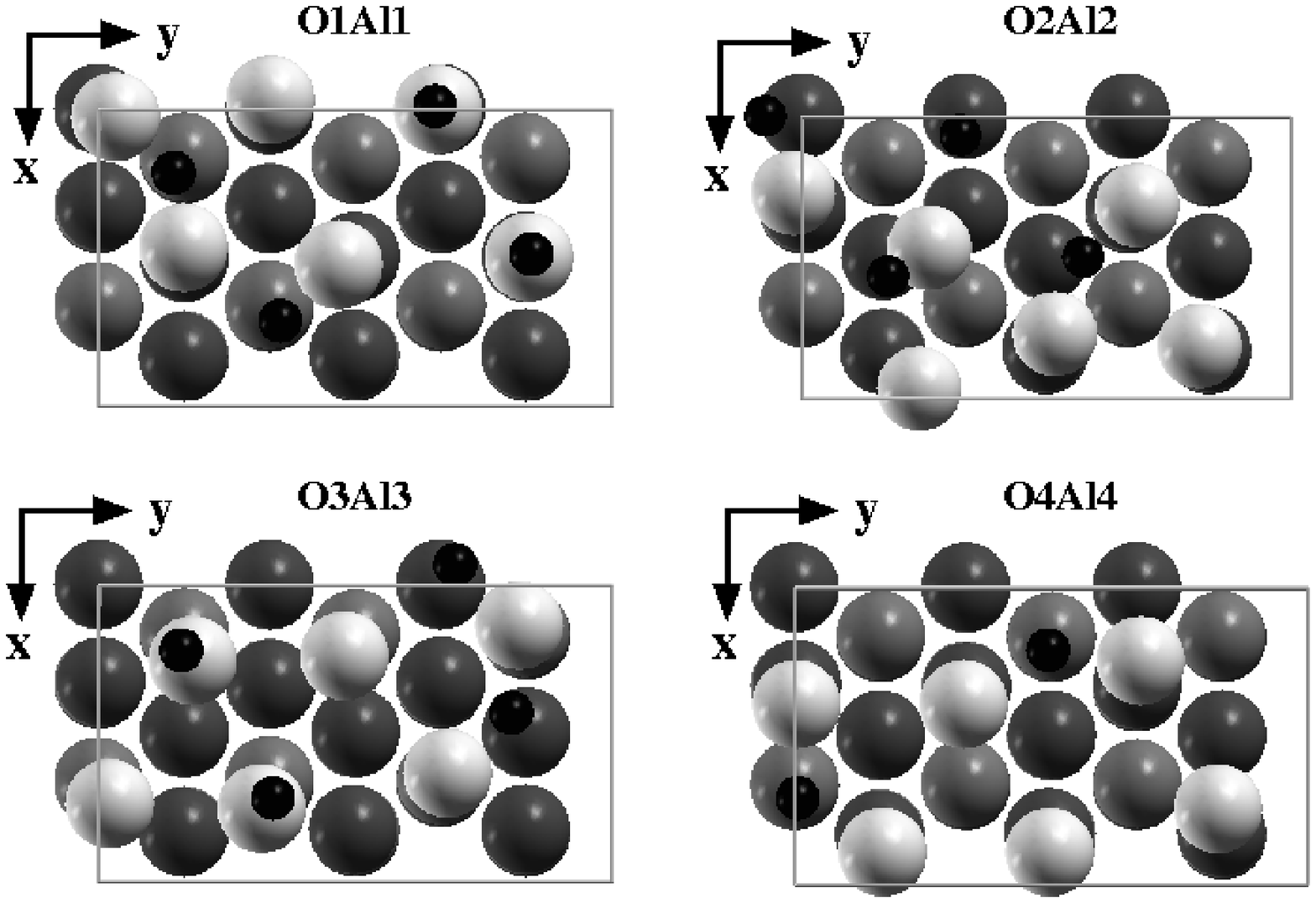,width=8.4cm}&
\epsfig{file=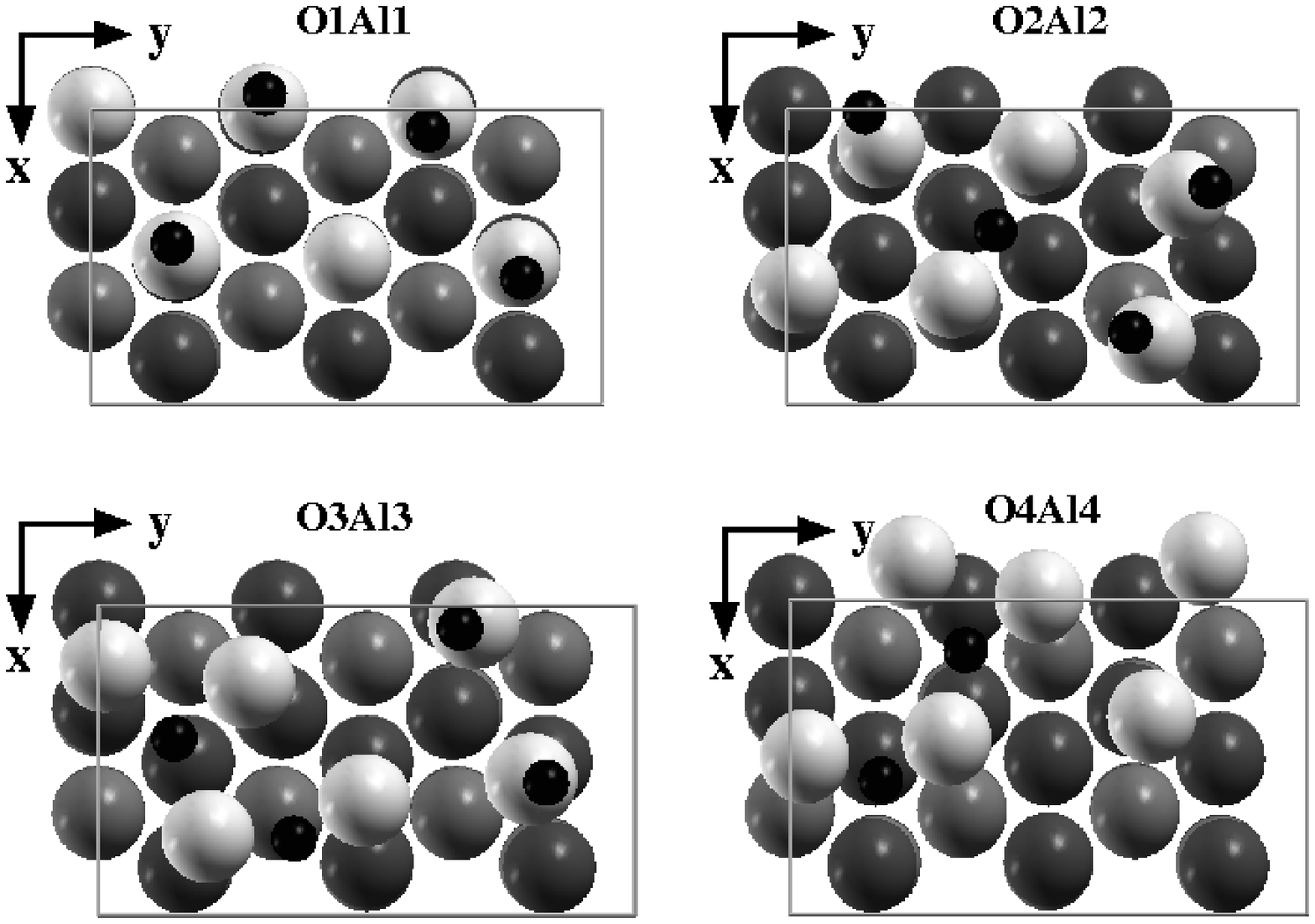,width=8.4cm}\\
\textbf{(a)}&\textbf{(b)}
\end{tabular}
\caption{
\label{fig:Al_metaIV}
Atomic structure of the potentially  metastable 
Al$_{14}$O$_{24}$ films. 
Color coding and notation 
are as in Fig.~\ref{fig:II-Al}.
The films are Al terminated
even after relaxations.
The Al coordinations are (without surface Al) 
(a)~$OT_{\downarrow}:OO:T_{\downarrow}O$,
(b)~$T_{\downarrow}T_{\downarrow}:T_{\downarrow}
(t_{\downarrow}o):T_{\downarrow}O$
($O$: octahedral, $T$: tetrahedral, 
$t$ single tetrahedral Al ion,
$o$ single octahedral Al ion,
the arrows indicate the direction in which the 
tetrahedra point).
}
\end{figure*}

\begin{figure}
\begin{tabular}{c}
\epsfig{file=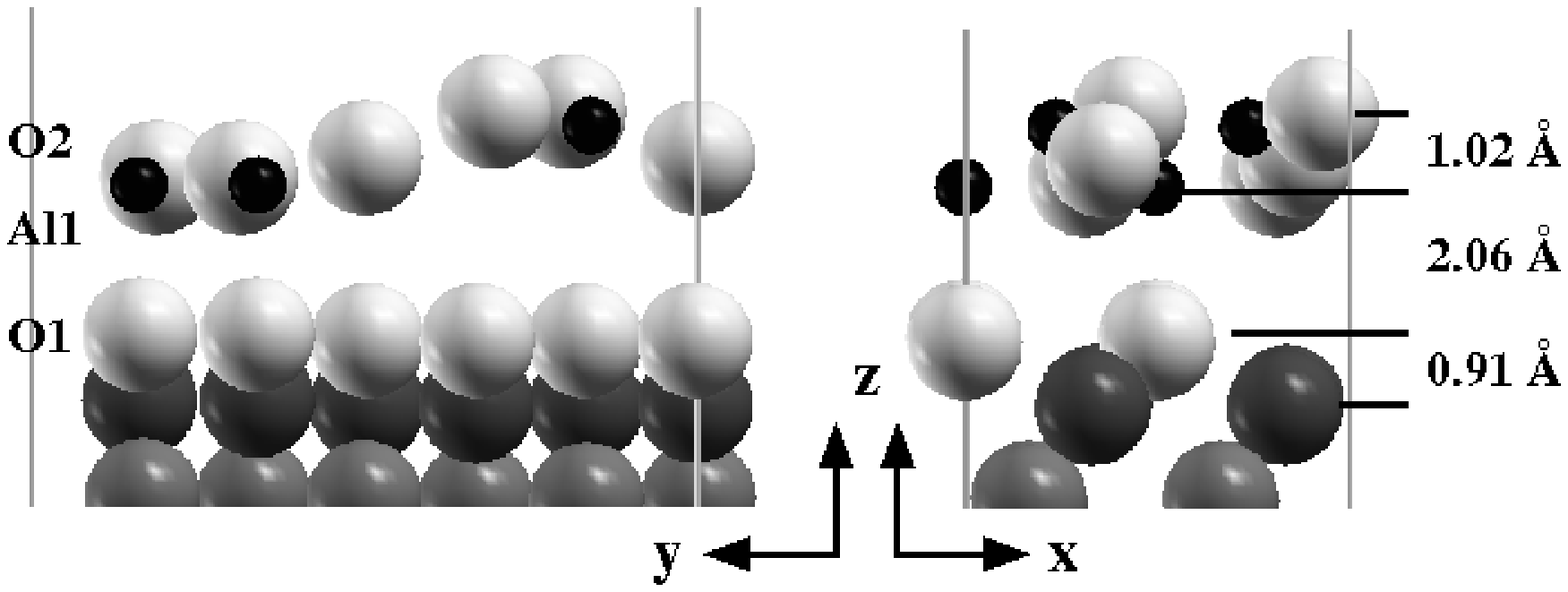,width=8.4cm}\\
\epsfig{file=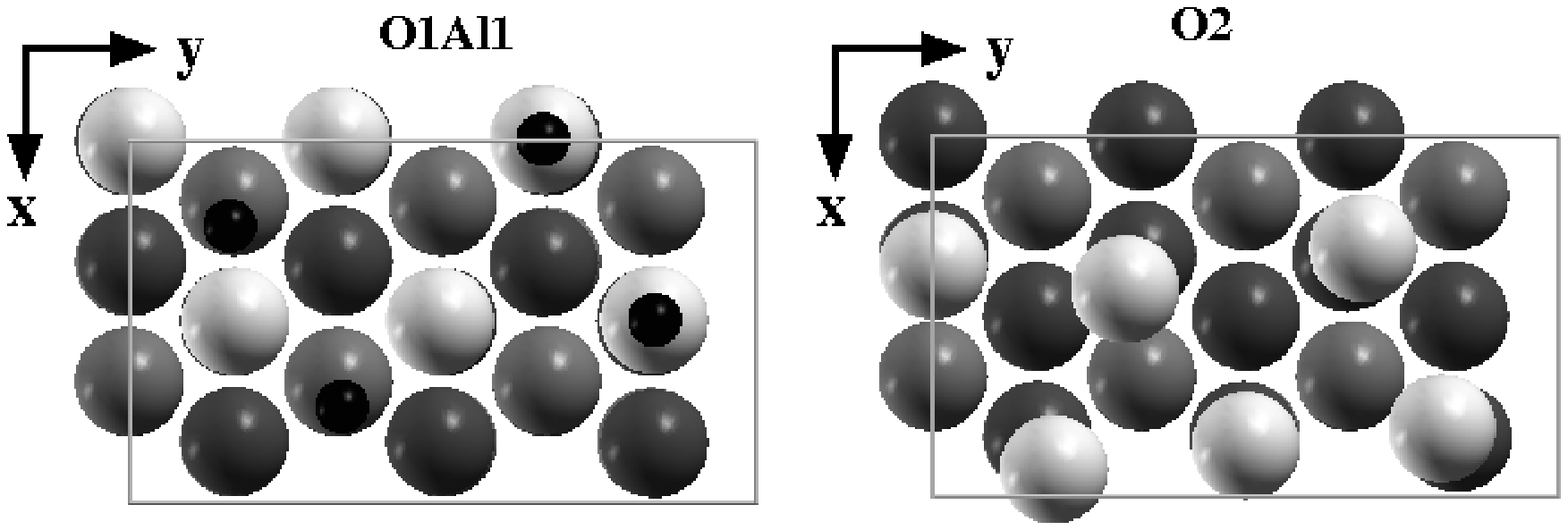,width=8.4cm}
\end{tabular}
\caption{
\label{fig:II-O}
Atomic structure of the stable 
Al$_4$O$_{12}$ 
film (color coding and notation as in Fig.~\ref{fig:II-Al}).
Note the large 
O--O interlayer distance, which forms an 
almost empty region in between the 
two O layers.
As a consequence, the TiC/alumina 
system separates into TiC/O/alumina.
The Al coordination is $T_{\downarrow}O$ 
($O$: octahedral, $T$: tetrahedral, 
the arrow indicates the direction in which the 
tetrahedra point). 
}
\end{figure}

\begin{figure}
\begin{tabular}{c}
\epsfig{file=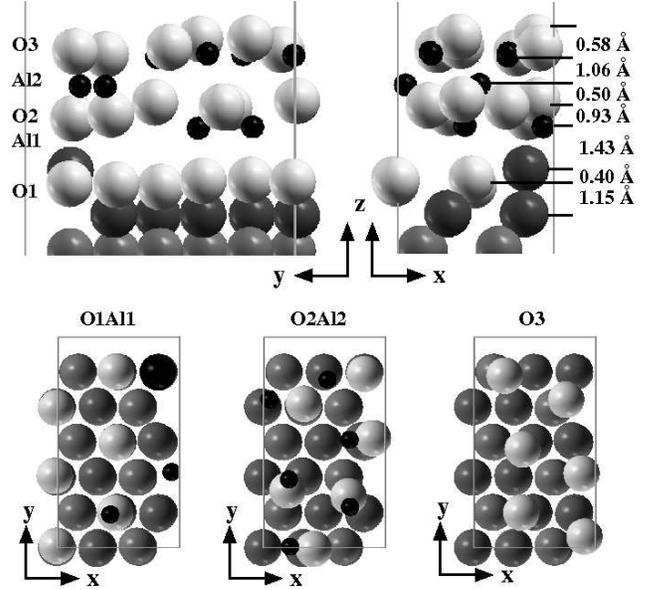,width=8.4cm}
\end{tabular}
\caption{
\label{fig:III-O}
Atomic structure of the stable 
Al$_8$O$_{18}$ 
film (color coding and notation as in Fig.~\ref{fig:II-Al}).
Note that one of 
interfacial Ti atoms 
has relaxed to a position 
slightly above the 
bottom O layer. 
In the top view on O1Al1 in
the lower panel,
this Ti impurity is indicated by 
the large black ball.
The Ti impurity may 
strengthen the TiC--alumina bond.
The Al coordination
is $T_{\downarrow}:OT_{\downarrow}T_{\downarrow}$
($O$: octahedral, $T$: tetrahedral, 
the arrows indicate the direction in which the 
tetrahedra point).
}
\end{figure}

\begin{figure}
\begin{tabular}{c}
\epsfig{file=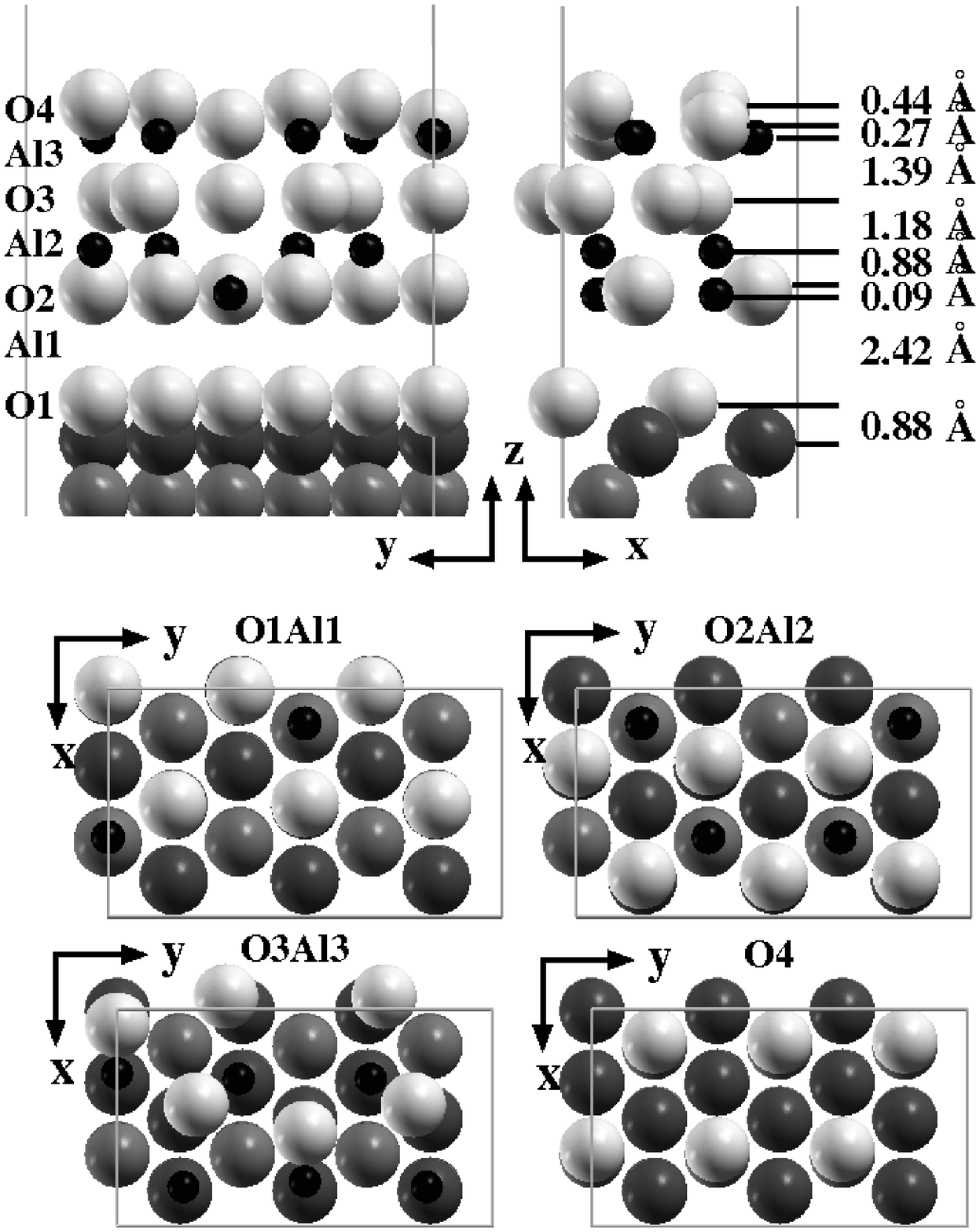,width=8.4cm}
\end{tabular}
\caption{
\label{fig:IV-O}
Atomic structure of the stable 
Al$_{12}$O$_{24}$
film (color coding and notation as in Fig.~\ref{fig:II-Al}).
Note again the large 
interlayer distance 
between the 
bottom two O layers. 
The absence of Al ions 
in between these O layers
strongly indicates 
that the TiC/alumina 
system separates into 
weakly bound TiC/O/alumina. 
The Al coordination
is $O:OO:OOO$, 
{\it i.e.}, purely octahedral. 
}
\end{figure}

\section{Results II: Atomic structure of stable and metastable films
\label{sec:Structure}}
In this section we analyze in detail the atomic
structure of the relaxed alumina films
with Al$_{4n-2}$O$_{6n}$ and Al$_{4n-4}$O$_{6n}$ stoichiometries 
that are found to be
stable and potentially metastable
in our implementation of the proposed method for structure search.
While a number of structural motifs that were included
in the initial candidate set are preserved,
also novel motifs that strongly deviate from 
those in the bulk phase are identified.
Both preserved and  novel motifs together,
give insight into structural motifs of more accurate thin-film candidates.

\subsection{Atomic structure of the  Al$_{4n-2}$O$_{6n}$ films} 
\setcounter{paragraph}{0}
\textit{Two-O-layer thick films -- Al$_6$O$_{12}$. }
Figure~\ref{fig:II-Al} shows the atomic structure of the
energetically most favorable Al$_6$O$_{12}$.
It corresponds to a close-packed
continuation of the TiC $ABC$ substrate stacking,
that is,
the alumina stacking is 
$Ab_{\alpha}b_{\beta}b_{\gamma}C$. 
All Al ions share the same atomic plane 
and are octahedrally ($O$) 
coordinated.  
The relaxed film is O terminated.  
Compared to TiC/O [O monolayer on TiC(111)], 
the Ti--O layer separation
is drastically increased 
($+0.5$~\AA).

The potentially metastable 
Al$_6$O$_{12}$ 
structures possess almost the same structure
as the 
energetically most favorable 
one.
They differ only 
by a slight displacement along the $z$ direction of
some of the Al ions.

\textit{Three-O-layer thick films -- Al$_{10}$O$_{18}$. }
Figure \ref{fig:III-Al} shows the atomic structure
of the energetically most favorable 
Al$_{10}$O$_{18}$ film. 
Potentially metastable configurations 
are displayed in Fig.~\ref{fig:Al_metaIII}.
In all cases, the Ti--O layer separation is 
shorter than in the  
energetically most favorable 
Al$_6$O$_{12}$ film,
but still considerably larger than in TiC/O 
($\sim+0.3$~\AA).
We  also note that in the 
stable Al$_{10}$O$_{18}$ film, 
two of the six O ions
in the 
bottom O layer are slightly
lifted off from the TiC substrate.
In the potentially metastable films,
no O ion is lifted off.

In all displayed films,
the surface Al pairs have relaxed below 
the terminating O layer,
so that the second Al layer consists of three Al 
pairs and the film is O terminated.

The stacking of O layers is 
approximately described by 
$ABA$, $AB(AC)_{\mtext{bridge}}$, and $ACA$
for the energetically most favorable film
and the two potentially metastable films respectively.
The order is  only approximate because 
a number of O ions are significantly distorted 
from ideal sites (as defined by the underlying TiC substrate).
They are often located in bridge or cusp sites.
This effect is most pronounced  in the third O layer 
in the first potentially metastable film,
which is entirely located in bridge sites.

Similarly, the Al ions often deviate from 
ideal sites so that their description
in terms of the bulk stacking labels
becomes cumbersome.
However, these distortions always
occur pairwise,
that is, Al pairs that are related 
by a bulk stacking label are dislocated symmetrically.
The candidate structures generally preserve this
symmetry of the motifs of the bulk phases.

The coordination of the Al ions 
is described as
$T_{\uparrow}T_{\downarrow}:T_{\downarrow}OO$,
$OO:T_{\downarrow}T_{\downarrow}T_{\downarrow}$
and $OO:OT_{\downarrow}T_{\downarrow}$,
for the energetically most favorable 
and the two metastable films respectively.
Here and in the following
$O$ denotes octahedrally coordinated Al pairs,
$T$ tetrahedrally coordinated pairs.
For tetrahedral coordination, 
$T_{\uparrow}$ means that the 
tetrahedra point along the TiC$[111]$ direction, 
away from the interface, 
whereas $T_{\downarrow}$ indicates
that they point 
towards the interface. 
Different Al layers are separated by '$:$'.

We notice that in all configurations,
there is a large number of tetrahedrally
coordinated Al ions
(40-60\%) and these can share the same atomic layer.
In particular a larger number of tetrahedrally coordinated Al
ions is favored.
Furthermore, the energetically most favorable film
contains  tetrahedrally coordinated Al ions
that share one layer and for which 
the tetrahedra point into opposite directions.

\textit{Four-O-layer thick films -- Al$_{14}$O$_{24}$. }
Figures~\ref{fig:IV-Al} and \ref{fig:Al_metaIV}
show the  atomic structure of the energetically most favorable
film and the two potentially metastable Al$_{14}$O$_{24}$  films, respectively.
The Ti--O 
layer separations are
comparable to those in the 
energetically favorable  Al$_{10}$O$_{18}$ films.
Also, in the 
stable Al$_{14}$O$_{24}$ film, 
two of the six O ions
in the bottom O layer are slightly
lifted off from the TiC substrate,
whereas this is not observed
in the potentially metastable films.
In the most favorable film,
the surface Al pairs have relaxed below 
the terminating O layer,
so that the second Al layer consists of three Al 
pairs and the film is O terminated.
However, both potentially metastable films
are Al terminated even after relaxations.

The stacking of O layers is more strongly distorted 
as in the case of Al$_{10}$O$_{18}$ films,
in particular in the most favorable film.
From the figures, we find the approximate O stacking sequences
$A(BC)_{\mtext{bridge}}(A_{\alpha}A_{\gamma}C_{\beta})B$
[most favorable film Fig.~\ref{fig:IV-Al}],
$ACBC$ [first potentially metastable, Fig.~\ref{fig:Al_metaIV}.(a)], and $ABAC$
[second potentially metastable film, Fig.~\ref{fig:Al_metaIV}.(b)],
where we use the the labeling (subscript) of the Al positions
also for O ions
and note that some of the ions are in fact dislocated from 
ideal sites.

The Al ions are distorted correspondingly. 
This distortion is again pairwise and symmetrically
for the energetically most favorbale and the first 
potentially metastable films.
For the second potentially metastable film,
this is not true.
Both on the surface and in the first and second layer below
the surface there are Al ions that have relaxed in a non-symmetric way.

The coordination of the Al ions is given by
$OT_{\downarrow}:T_{\downarrow}T_{\uparrow}:OT_{\downarrow}T_{\downarrow}$ (most favorable),
$OT_{\downarrow}:OO:OT_{\downarrow}$ (first potentially metastable), 
and $T_{\downarrow}T_{\downarrow}:oT_{\downarrow}t_{\downarrow}:ooT_{\downarrow}$ (second potentially metastable).
In the last sequence, the coordination of single ions 
that do not belong to a pair is denoted by small letters ($t$, $o$).
Also, the coordination of the surface Al ions is not given 
for the two potentially metastable films.

The result is similar to that for the Al$_{14}$O$_{24}$  films.
In general, a large  number of tetrahedrally coordinated Al ions 
is favored.
In the most favorable film 70\% of the Al ions
are tetrahedrally coordinated.
Furthermore,
there is a layer with purely tetrahedrally Al ions and 
tetrahedra pointing  into opposite directions
(second Al layer).
The first potentially metastable film possesses only 30\% tetrahedrally coordinated Al ions.
Inspection of the detailed stacking sequence,
$Ab_{\beta}a_{\gamma}Ca_{\alpha}a_{\beta}Bb_{\beta}a_{\gamma}Cb_{\alpha}$,
identifies this structure as a partial \Kalumina\ configuration 
with an orientation TiC$[111]$/$\kappa[00\bar{1}]$.

\subsection{Atomic structure of the Al$_{4n-4}$O$_{6n}$ films}
In general the potentially metastable Al$_{4n-4}$O$_{6n}$ films
(if present) possess the same atomic structure as the
the energetically most favorable film
but rotated by 180$^{\circ}$ around the TiC$[111]$ direction
[$B\leftrightarrow C, \beta(2)\leftrightarrow \gamma(3)$].
They are therefore not discussed in the following.

\setcounter{paragraph}{0}
\textit{Two-O-layer thick films -- Al$_4$O$_{12}$. }
Figure~\ref{fig:II-O} reports the calculated atomic
structure of the energetically most favorable Al$_4$O$_{12}$ film.
It is noticeable that the O--O
separation in the alumina is relatively large,
$d_{\mtext{O-O}}\sim2.6$~\AA\ on average,
and that the two Al pairs are not located
between the O layers
but are almost incorporated 
in the surface O layer,
which leads to a large
splitting of that layer.
At the same time, the 
Ti--O separation
is comparably small
and equals that in  TiC/O.

Hence, although predicted  to be stable in a
thermodynamical sense, structurally
this TiC/Al$_4$O$_{12}$
configuration separates into a TiC/O/Al$_4$O$_6$ system,
that is, a strongly bonded O
monolayer on the TiC substrate 
with a thin alumina overlayer on top.

The stacking of the O layers is $AC$,
and the coordination of the Al ions
approx $T_{\downarrow}O$.

\textit{Three-O-layer thick films -- Al$_8$O$_{18}$. }
Figure~\ref{fig:III-O} shows the atomic structure
of the energetically most favorable Al$_8$O$_{18}$ film.
The average O--O separation between  the 
bottom two O layers is $d_{\mtext{O-O}}\sim2.5$~\AA,
which is slightly shorter than the one in 
the stable Al$_4$O$_{12}$ film.
At the same time, the 
Ti--O separation is increased to $1.15$~\AA.

The stacking of the O layers is approximately $ACA$. 
In the middle O layer, the two O ions that should 
be located in $C_{\beta}$ are, however,
dislocated to cusp sites.
Furthermore the whole surface O layer is strongly
distorted from ideal sites.
The coordination sequence of the Al ions is
$T_{\downarrow}:OT_{\downarrow}T_{\downarrow}$.  

Only one of the 
original two Al pairs is left between the 
bottom two O 
layers after relaxation, 
the other pair 
moving in between the 
top two O layers. 
Interestingly, one of the interfacial
Ti atoms has left the 
Ti layer and relaxed slightly in between the 
bottom two O layers.
Again, structurally the
TiC/Al$_8$O$_{18}$ 
configuration 
appears as a partially decoupled TiC/O/Al$_8$O$_{12}$ system.
Here, however, the Ti impurity
above the bottom O layer 
may be a stabilizing factor.

\textit{Four-O-layer thick films -- Al$_{12}$O$_{24}$. }
In Fig.~\ref{fig:IV-O}, we show the atomic structure
of the stable Al$_{12}$O$_{24}$ film.  
The O stacking is $ACAB$ and hardly distorted.
All Al ions have octahedral coordination.
Thus, the present structure mixes 
the O stacking of bulk \Kalumina\
with the Al coordination of 
bulk \Aalumina.

The O--O separation  
$d_{\mtext{O-O}}\sim2.5$~\AA\ 
is again very large
and the Ti--O separation 
is TiC/O like.
Also, one of the  original 
two Al pairs in the 
bottom Al layer has 
relaxed upward 
through the 
middle Al layer and 
into the top layer.
The other Al pair 
of the bottom layer is 
after relaxation located only $0.1$~\AA\ below the 
middle O layer.
Consequently, also 
the TiC/Al$_{12}$O$_{24}$ configuration
can again be considered as a decoupled, weakly binding
TiC/O/Al$_{12}$O$_{18}$ system.


\section{Discussion \label{sec:Discussion}}

\subsection{Thin-film structure search method} 
\begin{table*}
\begin{ruledtabular}
\begin{tabular}{cll}
$n$ (\# of    & \mc{2}{c}{Favorable O stacking and Al coordination}\\
O layers) & Al$_{4n-2}$O$_{6n}$&Al$_{4n-4}$O$_{6n}$\\
     
\hline
2        & O$_A$ Al$^O$ Al$^O$ Al$^O$ O$_C$ 
         & O$_A$ Al$^O$ Al$^{T_{\downarrow}}$  O$_C$ \\
3        & O$_A$ Al$^{T_{\uparrow}}$ Al$^{T_{\downarrow}}$ O$_B$ Al$^O$ Al$^{T_{\downarrow}}$ Al$^O$ O$_A$
         & O$_A$ Al$^{T_{\downarrow}}$ O$_C$ Al$^O$  Al$^{T_{\downarrow}}$ Al$^{T_{\downarrow}}$ O$_B$       \\
         & O$_A$ Al$^O$ Al$^O$ O$_C$ Al$^{T_{\downarrow}}$ Al$^{T_{\downarrow}}$ Al$^{T_{\downarrow}}$ O$_A$ &      \\
         & O$_A$ Al$^O$ Al$^O$ O$_C$ Al$^O$ Al$^{T_{\downarrow}}$ Al$^{T_{\downarrow}}$  O$_A$     &  \\
4        & O$_A$ Al$^O$ Al$^{T_{\downarrow}}$ O$_B$ Al$^{T_{\uparrow}}$ Al$^{T_{\downarrow}}$ O$_A$ Al$^O$ Al$^{T_{\downarrow}}$ Al$^{T_{\downarrow}}$ O$_C$
         & O$_A$ Al$^O$ O$_C$ Al$^O$ Al$^O$ O$_A$ Al$^O$ Al$^O$ Al$^O$ O$_B$   \\
         & O$_A$ Al$^O$ Al$^{T_{\downarrow}}$ O$_C$ Al$^O$ Al$^O$ O$_B$ Al$^O$ Al$^{T_{\downarrow}}$ O$_C$ Al& 
\end{tabular}
\end{ruledtabular}
\caption{
\label{tab:FavorableStructuralMotifs}
Favorable structural motifs for relaxed films of different
thicknesses.
The label of the O layers indicate the approximate location of that layer
with respect to the substrate surface ($A$ = fcc, $B$ = hcp, and $C$ = top).
The label of the Al pairs indicate their approximate coordination 
($O$: octahedral, $T$: tetrahedral, 
the arrows indicate the direction in which the 
tetrahedra point).
}
\end{table*}

Although the relations between unrelaxed alumina structures 
and their energies after relaxations show some 
general trends, see  Sec.~\ref{sec:InterfacialOrientation},
several important exceptions occur. 
In particular, these exceptions can result into 
the energetically most favorable structure.

These exceptions illustrate a potential danger of applying
simple MC methods to the problem of finding the
stable thin-film oxide structures. 
An importance sampling of the thin-film configuration 
space based on a classification of the unrelaxed structures 
in terms of, for instance, alumina phase content, orientation, 
and/or O stacking may easily miss such exceptions. 

In our analysis of the relaxed atomic structure, we 
find that (apart from one possibly  metastable 
Al$_{12}$O$_{24}$ film) 
none of the stable and potentially metastable films 
shows a partial bulk alumina structure.
Nevertheless in all structures the Al ions
still obey parts of the bulk symmetry.
In particular, the positions of the Al pairs 
in the relaxed films are still pairwise related
by the mapping given in Fig.~\ref{fig:Bulk}.

The set of thin-film candidates does have new structure motifs
different from those found in the bulk phases. 
The thin-film relaxation causes differences
the coordination of the Al pairs
and/or in the stacking sequence of the O planes. 
For the  Al$_{4n-2}$O$_{6n}$ 
films we find that 
(i) layers with only tetrahedrally coordinated 
Al ions, $T_{\downarrow}T_{\uparrow}$ or $T_{\downarrow}T_{\downarrow}$, 
are energetically favorable; 
(ii) layers with only octahedrally coordinated Al ions 
are present in both stable and metastable films; 
(iii) Al layers with coordination $T_{\uparrow}T_{\uparrow}$ 
are not present. 
For the Al$_{4n-4}$O$_{6n}$ films we 
find essentially only Al coordinations of the 
types $OO$ and $OT_{\downarrow}$.  
The $T_{\downarrow}T_{\downarrow}O$ coordination 
of the Al layer directly below the surface O layer of the 
stable Al$_8$O$_{18}$ 
film should be considered as a surface effect.
The absence of  $OT_{\uparrow}$ 
is consistent with the fact
that the stable configurations
all derive from TiC/$\kappa[00\bar{1}]$ sequences.
However, the relaxed configurations are still not
conform with the bulk stacking.
In particular, 
for the stable Al$_{12}$O$_{24}$ film,
we observe only octahedrally coordinated
Al ions in combination with an  $ACAB$ stacking.

The finding of new structure motifs,
not explicitly included in the network 
of initial configurations,
implies a significant strength.
It shows that the proposed method
is not restricted to a sorting of  
the original candidate structures
in an energetic order,
but  that it is indeed capable to predict
energetically more favorable film geometries
than what strictly constitutes symmetries in partial 
bulk structures.

The identification of candidates for stable and
metastable thin-film structures with some novel structural
motifs also suggests how the candidate space
and the search could be broadened in a 
cost-efficient approach. 
A broadening of the network of
initial thin-film configurations 
can be made in the scope of structural elements.
It would include motifs 
found in the bulk $\alpha$- and $\kappa$-Al$_2$O$_3$ but
with a different weighting in choice of coordination for 
Al ions.  
In particular, the positions of the Al ions are still
related pairwise, which restricts 
any broadening of the network of necessary initial
thin-film configurations very significantly.
We find indications for a significant increase 
in preference for tetrahedral
coordination of Al ions in the thin-film candidates.
This motif is included in the initial thin-film network (which 
has many structures derived from $\kappa$-Al$_2$O$_3$).
However, a natural further refinement of the present
implementation of the proposed search method would be 
to include initial structures with a higher degree of
tetrahedral coordination of Al ions.
It is possible to cast this broadening 
of the initial network
into the framework of a genetic algorithm 
for identifying surface reconstructions.
\cite{ref:Chuang_GeneticAlg,ref:Sauer_GeneticAlg,ref:Rohrer_GeneticAlg}

We emphasize that a future, extended search for thin-film 
candidates is not expected to affect conclusions concerning
thermodynamical stability of the various classes of ultra-thin
alumina films.
Since the slopes of $\Gamma$ in 
Figs.~\ref{fig:GibbsEquilibrium}
will remain unchanged,
finding possibly  energetically more favorable structures
in the two relevant stoichiometry classes
will only resize the regions in which
the different stoichiometries are stabilized.
To make the Al$_{4n-4}$O$_{6n}$ films generally unstable
in comparison to Al$_{4n-2}$O$_{6n}$ films,
the truly stable Al$_{4n-2}$O$_{6n}$ configuration
needs to gain at least $\sim 10$~eV compared
to the stable Al$_{4n-4}$O$_{6n}$ found here.

\subsection{Note on the stability of CVD TiC/alumina wear-resistant coatings}
We emphasize that our results on the thermodynamical stability
of thin-film alumina on TiC are critically based on the assumption
of thermal equilibrium between the films and an O$_2$ environment.
The finding that the  preferred structure is described
as a TiC/O/Al$_{4(n-1)}$O$_{6(n-1)}$ system (after relaxations)
with a weak binding \cite{ref:QuantitativeBinding}
between TiC/O and Al$_{4(n-1)}$O$_{6(n-1)}$
is not in contradiction with the required exceptionally strong binding
in a wear-resistant coating application.
It rather shows that equilibrium between O$_2$ and oxygen
in the alumina during CVD growth of alumina 
is not reached.
At environmental conditions relevant for CVD growth,
as will be be discussed in a forthcoming paper,
\cite{ref:CVD_Thermodynamics}
we find that it is instead the Al terminated films 
that are stabilized.

\section{Conclusions\label{sec:Summary}}
We present a method to sample the
configuration space of possible 
thin-film structures of complex oxides
on a substrate.
A well-defined network of initial configurations for
promising thin-film candidates can be designed
from the oxide bulk structure.
\textit{Ab initio} calculations of 
relaxation deformations provide candidates for 
thin films as a function of stoichiometry and 
oxygen-layer thickness.

The method has been illustrated 
for TiC/thin-film alumina,
where experimental evidence \cite{ref:TiX-Al2O3_Coatings,ref:Al2O3TiX_TEMGrowth}
can be used to reduce 
the network of initial thin films 
to contain structural motifs defined by 
bulk $\alpha$- and \Kalumina.
Based on this assumption,
we have determined 
structural elements in and candidates for
the energetically most favorable 
(stable or potentially metastable)
TiC/thin-film alumina  configurations
for three thicknesses and three stoichiometry classes.

Our method for {\it ab initio} 
search and study of thin-film structures
has predictive power and 
provides detailed insight 
into the nature and atomic structure
of thin-film alumina on TiC. 
The structures that are predicted by our method
differ in their motifs heavily 
from motifs of the bulk structures,
in particular in terms of the Al coordination. 
In principle,
this warns that the present implementation of the search
may not yet be complete and 
that we cannot make an 
authoritative prediction of the stable thin-film 
alumina structure; 
we can at present only identify 
key structural elements.
More importantly, this finding of 
additional favorable motifs documents 
predictive power.
It shows that the search method can identify
candidates with a nature that is not explicitly included
in the network of initial configurations.

The different stoichiometry classes 
have been compared by means of Gibbs free energies.
Assuming equilibrium with an O$_2$ environment,
we find that for the considered thicknesses of two, three, or four 
O layers (corresponding to $n = 2$, $3$, or $4$, respectively) 
the stable films are either
those with Al$_{4n-4}$O$_{6n}$ stoichiometry 
(for medium to high O chemical potentials) 
or  those with Al$_{4n-2}$O$_{6n}$ stoichiometry 
(for very low O chemical potentials). 
The films with Al$_{4n}$O$_{6n}$ stoichiometry 
are never stabilized.

\section*{Acknowledgments}

The authors thank Sead Canovic and Mats Halvarsson for useful discussions.
Support from the Swedish National Graduate School in Materials Science,
from the Swedish Foundation for Strategic Research (SSF) through ATOMICS, 
from the Swedish Research Council (VR), and from the Swedish National 
Infrastructure for Computing (SNIC) are gratefully acknowledged.




\begin{references}

\bibitem{ref:AlxOyStructures}
$\alpha$-,$\gamma$-, $\delta$-, $\theta$-, \Kalumina, \ldots,
see \textit{e.g.}
I.\ Levin and D.\ Brandon,
J.\ Am.\ Ceram.\ Soc.\ \textbf{81}, 1995 (1998).

\bibitem{ref:TixOyStructures}
rutile-, anatase-, brookite, and columbite ($\alpha$-PbO$_2$) TiO$2$,
and Ti$_2$O$_3$,
see \textit{e.g.}
J.\ Haines and J.\ M.\ Leger,
Physica B, \textbf{192},  233 (1993);
J.\ K.\ Dewhurst and J.\ E.\ Lowther, 
Phys. Rev. B 54, R3673 (1996).


\bibitem{ref:VxOyStructures}
VO$_x$ (rocksalt), VO$_2$ (rutile), V$_2$O$_3$ (corundum), V$_2$O$_5$ (orthorhombic),  
see \textit{e.g.}
S.\ Surnev, M.~G.\ Ramsey and F.~P.\ Netze,
Prog.\ Surf.\ Sci.\ \textbf{73}, 117 (2003).


\bibitem{ref:HfxOyStructures}
cubic, tetragonal, and monoclinic modifications of HfO$_2$, see \textit{e.g.}
J.\ Wang, H.\ P.\ Li and R.\ Stevens,
J. Mat. Sci. \textbf{27}, 5397 (1992).



\bibitem{ref:Stierle_NiAl-Alumina}
A.\ Stierle \etal, 
Science \textbf{303}, 1652 (2004).

\bibitem{ref:Kresse_NiAl-Alumina}
G.\ Kresse \etal, 
Science \textbf{308}, 1440 (2005).

\bibitem{ref:TEM}
S.\ Canovic \etal,
Surf.\ Coat. Technol.\ \textbf{202}, 522 (2007).


\bibitem{ref:SEM}
M.\ Halvarsson \etal, 
J.\ Phys.:\ Conf.\ Ser.\ \textbf{126}, 012075 (2008) .


\bibitem{ref:TiX-Al2O3_Coatings}
M.\ Halvarsson, H.\ Nord\'en, and S.\ Vuorinen, 
Surf.\ Coat.\ Technol.\  \textbf{61}, {177} (1993).


\bibitem{ref:Al2O3TiX_TEMGrowth}
M.\ Halvarsson, J.E.\ Trancik, and S.\ Ruppi, 
Int.\ J.\ Refract.\ Met.\ Hard Mater.\ \textbf{23}, 32 (2006).
%


\bibitem{ref:Rohrer_CoarseGrained}
J.\ Rohrer \etal, 
J.\ Phys.:\ Conf.\ Ser.\ \textbf{100}, 082010 (2008) 
%
\bibitem{ref:NoteOnComplexity}
Both alumina phases, in particular \Kalumina, yield a huge number of possible
thin-film configurations. 
The primitive unit cell of \Kalumina,
with its $ABAC$ stacking of O planes
along the $[001]$ direction,
see also Fig.~\ref{fig:Bulk},
allows for 
$\tiny{
\left(
\begin{array}{c}
18\\
4
\end{array}
\right)}=3060
$
combinatorially
possible distributions of the
four Al ions
within each atomic plane.
Use of symmetry and 
electrostatics arguments
(for example, no occupation of 
nearest-neighbor sites for Al), 
reduces this number to 222.
However, it is clear that for thin films
of a few atomic layers,
the number of possible
atomic structures increases 
rapidly.


\bibitem{ref:MD_general}
R.\ Car and M.\ Parrinello,
Phys.\ Rev.\ Lett.\ \textbf{55}, 2471 (1985).


\bibitem{ref:Kaxiras_Thermodynamics}
E.\ Kaxiras \etal,
Phys. Rev. B \textit{35}, 9625 (1987).


\bibitem{ref:Finnis_Thermodynamics}
I.\ G.\ Batyrev, A.\ Alavi, and M.\ W.\ Finnis,
Phys. Rev. B \textbf{62}, 4698 (2000).


\bibitem{ref:Scheffler_RuO2}
K.\ Reuter and M.\ Scheffler,
Phys.\ Rev.\ B \textbf{65}, 035406 (2001).


\bibitem{ref:alphaStructure}
L.\ Pauling and S.\ B.\ Hendricks, 
J.\ Am.\ Chem.\ Soc.\ 
\textbf{47}, 781 (1925);
M.\ L.\ Kronberg, Acta Metall. \textbf{5}, 507 (1957);
W.\ E.\ Lee and K.\ P.\ D.\ Lagerlof,
J.\ Electron.\ Microc.\ Tech.\
\textbf{2}, 247 (1985).

\bibitem{ref:Yashar_KappaBulk}
Y.\ Yourdshahyan \etal, 
J.\ Am.\ Ceram.\ Soc. \textbf{82},  1365 (1999).


\bibitem{ref:Carlo_Surfaces}
C.\ Ruberto, Y.\ Yourdshahyan, and B.\ I.\ Lundqvist,
Phys.\ Rev.\ B  \textbf{67}, 195412 (2003).


\bibitem{ref:Alpha_latticeParameterExp}
W.\ E.\ Lee and K.\ P.\ D.\ Lagerlof,
J.\ Electron.\ Microc.\ Tech.\
\textbf{2}, 247 (1985).

\bibitem{ref:Kappa_latticeParameterExp}
M.\ Halvarsson, V.\ Langer, S.\ Vuorinen,
Surf.\ Coat.\ Technol.\
\textbf{76-77}, 358 (1995).


\bibitem{ref:Halvarsson_alpha}
M.\ Halvarsson,  Ph.\ D.\ Thesis, Chalmers University of Technology (1994).

\bibitem{ref:AdsorptionOnTiX}
C.\ Ruberto and B.\ I.\ Lundqvist, 
Phys.\ Rev.\ B \textbf{75}, 235438 (2007);
A.\ Vojvodic, C.\ Ruberto, and B.\ I.\ Lundqvist, 
Surf.\ Sci.\  \textbf{600},  {3619} (2006).

\bibitem{ref:TiCLatticeExp}
A.\ Dunand, H.\ D.\ Flack and K.\ Yvon, 
Phys. Rev. B \textbf{31}, 2299 (1985).



\bibitem{ref:TiC_TerminationExp}
C.\ Oshima \etal 
J.\ Less-Common Met.\ 82, p. 69 (1981).



\bibitem{ref:Sead}
S.\ Canovic \etal, 
Surf.\ Coat.\ Technol.\ \textbf{202} 522 (2007).


\bibitem{ref:Dacapo}
dacapo, \textit{https://wiki.fysik.dtu.dk/dacapo}.


\bibitem{ref:PSP}
D.\ Vanderbilt, 
Phys.\ Rev.\ B \textbf{41}, 7892 (1990).


\bibitem{ref:PW91}
J.\ P.\ Perdew \etal,
Phys.\ Rev.\ B \textbf{46}, {6671} (1992).


\bibitem{ref:MP}
H.\ J.\ Monkhorst and J.\ D.\ Pack,
Phys.\ Rev.\ B \textbf{13}, 5188 (1976).





\bibitem{ref:Carlo_PhD}
C.\ Ruberto, 
Ph.\ D.\ Thesis, Chalmers University of Technology (2001).





\bibitem{ref:alphaSurfaces_Thermodynamics}
X.-G.\ Wang, A.\ Chaka, and M.\ Scheffler,
Phys.\ Rev.\ Lett.\  \textbf{84}, 3650 (2000);
A.\ Marmier and S.\ C.\ Parker, 
Phys.\ Rev.\ B  \textbf{69}, 115409 (2004).





\bibitem{ref:Zhang_AluminaInterface}
W.\ Zhang, J.\ R.\ Smith, and X.-G.\ Wang, 
Phys.\ Rev.\ B  \textbf{70}, 024103 (2004).




\bibitem{ref:GammaVib}
Note that, even in the absense of $\delta$
vibrational contributions are of minor relevance:
First, our results found by excluding $\Gamma^{\mtext{vib}}$
show that the in differences $\Gamma$ for different
stoichiometric film compositions
are of the order of $5-10$~eV in the 
largest range of most interesting
region of the O chemical potential.
Second, although small regions were these differences 
become of the order of $\Gamma^{\mtext{vib}}$ exist
we have to keep in mind that it is not the absolute
value of $\Gamma^{\mtext{vib}}$ but
rather the differences in $\Gamma^{\mtext{vib}}$
for different surface terminations
that determine the stability.
These can be expected to be considerably smaller
than the absolute value of $\Gamma^{\mtext{vib}}$.
Hence, the only effect of neglecting
vibrational contributions is a small uncertainty
in the value of $\mu_{\mtext{O}}$ which 
divides regions where different stoichiometries
are stable.



\bibitem{ref:JANAF}
NIST-JANAF Thermochemical Tables $4^{\mtext{th}}$ Ed., 
Malcom W. Chase Jr. (1998).






\bibitem{ref:GammaWithDelta}
By including  a thickness and stoichiometry dependent value 
of $\delta$ we find that the Al$_{4n-4}$O$_{6n}$ films are still stabilized
down to $\mu_{\mtext{O}}\geq -2$ to $-2.5$~eV,
where the higher value applies for the thickest
and the lower for the thinnest films.
An O chemical potential of 
$\Delta\mu_{\mtext{O}}\geq -2$~eV,
is reached for considerably higher O$_2$ pressures 
(\textit{e.g.} $T\sim 1300$~K, $p_{\mtext{O}_2}\sim10^{-4}$~bar).
However, we note that the estimate of $\delta$ for 
the thicker films may be to large,
so that the resulting value of $\Gamma$
is too low and the value of the O chemical potential
$\Delta\mu_{\mtext{O}}\geq -2$~eV is too high.
In any case, 
at not too high temperatures,
and not too low pressures 
the Al$_{4n-4}$O$_{6n}$ stoichiometries
will always be stabilized.





\bibitem{ref:Chuang_GeneticAlg}
F.\ C.\ Chuang \etal,
Surf.\ Sci.\ \textbf{573}, L375 (2004).


\bibitem{ref:Sauer_GeneticAlg}
M.\ Sierka \etal,
J.\ Chem.\ Phys.\ \textbf{126}, 234710 (2007).


\bibitem{ref:Rohrer_GeneticAlg}
J.\ Rohrer and P.\ Hyldgaard,
unpublished.


\bibitem{ref:QuantitativeBinding}
This statement is based on geometrical grounds
(large separations between 
TiC/O and the alumina).
A more quantitative discussion of the anchoring
of thin  Al$_{4(n-1)}$O$_{6(n-1)}$ films on TiC/O
and Al$_{4n-2}$O$_{6n}$ films on TiC
will be given elsewhere.


\bibitem{ref:CVD_Thermodynamics}
J.\ Rohrer, C.\ Ruberto, and P.\ Hyldgaard,
unpublished.









\end{references}
\end{document}